\documentclass[12pt]{article
}

\usepackage{graphics} 
\usepackage{epsfig}
\usepackage{amssymb}
\usepackage{amsmath}
\usepackage{amsthm}
\usepackage{amstext}
\usepackage{graphicx}
\usepackage{dcolumn}
\usepackage{bm}

\begin{document}
\DeclareGraphicsExtensions{.jpg,.pdf,.mps,.png,} 
\noindent{\Large \bf Electronic structure and transport in graphene: quasi-relativistic
Dirac -- Hartry -- Fock self-consistent field approximation}

\noindent{\large {\it  H. V.~Grushevskaya and G. G.~Krylov}\\
E-mail: grushevskaja@bsu.by\\
Physics Department, Belarusian State University,
     4 Nezalezhnasti Ave., 220030 Minsk, BELARUS}%

\section{Introduction}
Graphene and graphene-like materials are considered today as a
prominent candidates to be used in new devices with functionality
based on quantum effects and(or) spin-dependent  phenomena in
low-dimensional systems. Technically, the main obstacle to such
devices implementation is the lack of methods that provide
minimizatioon of distortion of these material unique properties in
bulk nanoheterostructures.
Theoretical approaches and computer simulation play an important role in
systematic search of this kind of  nanostructures
for nanoelectronic applications.

However, significant part of all theoretical consideration of graphene-like
materials as well as  modern model representations of charge transport in these systems
on  pseudo-Dirac massless fermion model, originally based on tight binding approximation and applied to the
description of graphite
\cite{Wallace,
Semenoff,
Saito,
Reich
}, which is a bulk material.

According to this approach
\cite{Semenoff},
$ \pi_z$-electrons
in graphene are massless fermion type quasiparticle excitations moving
with the Fermi velocity. The approach has been seriously developed and successfully applied to
a number of experimental situation, some related review papers on the topics and further references are in \cite{Neto,Peres}.
There are few known key points where (at our knowledge) one could expect the necessity of some generalized consideration.
The first one is the cyclotron mass dependence upon the carriers concentration.
%
Due to weakness of the signal, modern experimental techniques can register
cyclotron mass of charge carriers which is just a little smaller than 0.02
of a free electron mass
\cite{Novoselov,
Zhang,
Deacon,
Jiang
}.
Assessments are absent  whether this mechanism of conductivity
prevails in the region of very small values  of charge carriers concentration.

The another point is the experimentally observable carrier asymmetry in graphene.
According to modern theoretical concepts the bands for pseudo-relativistic
electrons and holes in graphene  must be symmetrical.
With this in mind in the paper
\cite{Rojas-Cuervo}  
in the generalized gradient approximation
there were modelled the hexagonal
Si and
Ge, with the same structure as in  graphene.
But the electrons and holes bands
near the Dirac points
$K$ in  the Brillouin zone
turned out to be strongly asymmetric ones for both cases.
Firstly,
a Dirac cone deformation takes place
far away from a circular shape, as for the second,
the Dirac velocities for the valent and conduction bands are different
\cite{Rojas-Cuervo}.
Therefore, one can  assume the existence of some asymmetry in the
behavior
of pseudo-relativistic electrons and holes of the graphene as well.
Since the value of the asymmetry seems to be very small, for its
experimental observation one should use a highly sensitive method,
such e.g., as based on the measurement of noises \cite{Altshuler}.
For graphene, such a method is based on  large amplitudes
of  non-universal  fluctuations of charge carriers current in the
form of nonmonotonic $1/f$ noise in the crossover region of the
scattering
\cite{Rossi}.
At high charge densities, the contribution to the resistance of clean
graphene basically gives the scattering of charge carriers on
long-range impurities at ordinary
(symplectic) diffusion.
Regime of pseudo-diffusion with a charge carriers scattering  on
short-range impurities is realized in the vicinity of the Dirac
points of the Brillouin zone.
In the papers 
\cite{Rahman,
Guikema} 
measurements of quantum interference noise in a crossover
between a pseudodiffusive and symplectic regime
and magnetoresistance measurements in
graphene p-n junctions
have been performed,
which established asymmetric behaviour of pseudo-relativistic electrons and holes
based on asymmetric form of non-monotonic dependence of
noise   and magnetoresistance.
And the third known key point needed theoretical explanation is the replicas existence.
A weakly interacting epitaxial graphene on the surface of  Ir(111)
has a non-distorted hexagonal symmetry due to the weakness of the
interaction with the substrate in temperature range up to a room temperature.
Therefore in
ARPES (angle-resolved  photo-electron spectroscopy) spectra,
the perturbation of the band structure is manifested in the form of replica
of the inverted Dirac cone and mini-gaps in places of quasi-crossings of
replicas and the Dirac cone \cite{Pletikosi}.
Authors of  paper
\cite{Pletikosi}
propose replicas existence explanation such that
replicas are produced only in the areas of convergence of C  and Ir atoms,
that explains the weak intensity of  photoelectron emission replicas
and brightness of the main cone.
However, besides different intensity, the asymmetry of photoelectron
emission spectra also manifests itself in the fact that
 maxima from  the replicas are below then  zero maximum
($ E-E_F \approx 0 $) of Dirac cone in the ARPES spectrum at the same
angle of incidence of photons and with are equal to maxima of ARPES spectrum of
replicas oppositely arranged on the hexagon.
And conversely, if the corners of oppositely disposed replicas are
in the neighborhood $ E-E_F \approx 0 $, the Dirac cone in the
ARPES spectrum is located lower. The above described is possible if
axes of the Dirac cone and its replicas are not parallel.
It means that the top of the replica does not correspond to the corners of the
 hexagonal mini Brillouin zone, centered at the Dirac cone corner
for the epitaxial graphene on the surface of  Ir(111).
It has been also demonstrated that epitaxial graphene on SiC(0001) holds  a hexagonal
mini Brillouin zone
near the Dirac points
\cite{Zhou,
Gweon}.

And at last, a bit more philosophical but also important comment. The majority of modern software for
{\it ab initio} band structure simulations uses models being some variant of the Dirac equation or at least
take into account the  known relativistic corrections
to the Schr\"odinger equation  when attacking the problems.
The quasi-Dirac massless fermion approach based purely on tight binding non-relativistic Hamiltonian seems to be
oversimplified and hardly extendable.

With the goal to investigate the balance of
exchange and correlation interactions
 the {\it ab initio} band structure simulations
of loosely-packed solids(it was used the developed generalization of the LMTO method)
have been performed and demonstrated that the strong exchange leads
to  appearance of an energy gap in the spectrum whereas
 strong correlation interactions leads to tightening of this
gap
\cite{Mysolid state 1998}.
The spin-unpolarized  {\it ab initio} simulations of partial electron
densities of  two-dimensional graphite
have shown that the material is a semiconductor.
Interlayer correlations  tighten the energy gap that results in
semi-metal behaviour of three-dimensional graphite
\cite{Mysolid state 1998}.
This means that in the absence of correlation holes, the correlation
interaction
in a monoatomic carbon layer (monolayer) is weak in comparison with the
exchange.
This theoretical prediction for spin-nonpolarized graphene
were confirmed  experimentally in
\cite{Zhou,
Gweon},
where it was demonstrated
a bandgap  in bilayer graphene
on SiC(0001) and its
diminishing up to vanishing
in multilayer graphene.

By the way, in a
monolayer graphene on SiC(0001)
one observes the dispersion of the
Dirac cone apexes
\cite{Zhou,
Gweon}.

Experimentally manufactured quasi-two-dimensional systems, such as
graphene, carbon armchair nanotubes and ribbons
as well as some types of zigzag carbon nanotubes manifest metallic
properties (see, for example, 
\cite{ Neto,
Thomsen}).
In this regard, there are discrepancies between theoretical
predictions and experimental data.

Therefore, based on the results of {\it ab initio}
spin-unpolarized simulations of two-dimensional and three-dimensional
graphite \cite{Mysolid state 1998},
we can make the following assumption.
Carbon low-dimensional systems having the properties  similar to
graphite-like materials should possess  spin-polarized
electronic states with the correlation holes (a magnetic ordering).
Enhancing of correlation interaction due to  correlation-hole
contribution leads to  tightening of the energy gap and, as a
consequence, the emergence of semi-metal conductivity.

The approach we use has been developed earlier
in \cite{myKazan} and applied for graphene-like material in \cite{myNPCS2013}.

The goal of this chapter is to represent a Dirac -- Hartree -- Fock
self-consistent field quasirelativistic approximation for
 quasi-two-dimensional systems and to describe the origin of
 asymmetry of electron -- correlation hole carriers
in graphene-like materials.

\section{ Graphene model bands with correlation holes}
Single atomic layer  of carbon atoms (two-dimensional
graphite) is called  monolayer graphene.
Its hexagonal structure can be represented by two triangular
sublattices A, B \cite{Kelly}.
The primitive unit cell of the graphene contains two carbon atoms
C$_A$ and C$_B$
belonging to the sublattices A and B respectively.

The carbon atom has four valent electrons s, p$_x$, p$_y$, p$_z$.
Electrons s, p$ _x $, p$ _y $ are  hybridized in the plane of the
monolayer,
p$_z$-electron orbitals  form a half filled band of $ \pi $-electron
orbitals on a hexagonal lattice.

Let electrons with spin $ + \sigma $ "down" \ ("up") are placed on the
sublattice $ A \ (B) $, and the electrons with spin $ - \sigma $ "up" \
("down") -- on the sublattice $ B \ (A) $, as shown in Fig.~
1.
\begin{figure}[hbt]\label{fig1}
\includegraphics[width=8.cm,height=6.6cm,angle=0]{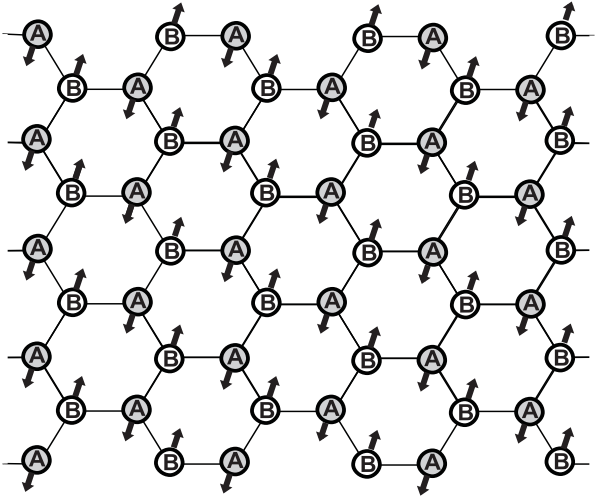}
\caption{Hexagonal  lattice of carbon monolayer with spin ordering
sublattices $A, \ B  $.  }
\end{figure}
With such a symmetry of the problem, all the relevant bands of
sublattices are half-filled and are formed
due to correlation holes.
In the representation of secondary quantization  and  Hartree -- Fock self-consistent  field
approximation, when not accounting for the
 electron density fluctuation correlation,
the hole energy $\epsilon (k)$  is simply added  to the
electron energy $\epsilon _m (0)$ \cite{Krylova monography2}:
\begin{eqnarray}
\left[
H(\vec r)+\hat V^{sc}(kr)-\hat \Sigma ^x (kr)
 \right]\psi_m (kr)
 = \left(\epsilon _m (0)- \sum_{j=1}^n
 \hat\epsilon^\dagger P_j
\right)\psi_m (kr), \\
\hat\epsilon^\dagger = \epsilon (k) \hat I
\end{eqnarray}
because the sum $\sum_j P_j$ of projection operators $P_j$ in in parentheses equals to
the identity operator $\hat I$: $\sum_j P_j = \hat I$.

We denote spinor wave functions of the valent electrons of graphene  as
$\widehat{\chi^{\dagger}_{\sigma}}(\vec r_A)\left|0,+ \sigma\right>$
and
$\widehat{\chi ^\dagger_{-\sigma}}(\vec r_B)\left|0,- \sigma\right>$.
From Fig.~
1 it follows that the spinor quantum fields
$ \widehat{\chi ^ {\dagger} _ {\sigma}} (\vec r_A) $ and
$ \widehat {\chi ^ \dagger _ {- \sigma}} (\vec r_B ) $
are transformed into each other under the mirror reflection $ A \to B $.
Therefore, a quasi-particle excitation in the proposed model
of graphene is a pair of an electron and a correlation hole.
As an electron-hole pairs at the same time represent
themselves their proper antiparticle,
the wave functions belong to the space of Majorana bispinors $ \Psi '$, and
upper and lower spin components $ \psi' $, $ \dot {\psi} '$
are transformed via different representations of the Lorenz group
\begin{eqnarray}
\Psi ' =\left(
\begin{array}{c}
\psi '_\sigma \\[0.1mm]
\dot{\psi} '_{-\sigma }
\end{array}
\right)=\left(
\begin{array}{c}
e^{{ \kappa \over 2}\vec \sigma \cdot \vec n}\psi _{\sigma }\\[0.1mm]
e^{{ \kappa \over 2}(-\vec \sigma )\cdot \vec
n}\dot{\psi}_{-\sigma }
\end{array}
\right). \label{majiran-neutrino-bispinor}
\end{eqnarray}
It means that
$\widehat {\chi ^{\dagger}_{\sigma}} (\vec r_A)\left|0,+ \sigma\right>$
behaves as a component
$\psi
_{\sigma }$, 
and $\widehat {\chi ^\dagger _{-\sigma}}(\vec
r_B)\left|0,- \sigma\right>$ -- 
as a component
$\dot{\psi}
_{-\sigma } $ of bispinor
(\ref{majiran-neutrino-bispinor}).
Using the expression (\ref{majiran-neutrino-bispinor}) and properties
of these operators:
$$
\widehat {\chi ^{\dagger}_{-\sigma_{_A} }}(\vec r)
\left|0,+ \sigma\right> \stackrel{def}{=\!=} 0,\
\widehat {\chi ^\dagger _{+\sigma_{_B}}}(\vec r)\left|0,- \sigma\right>
\stackrel{def}{=\!=} 0
$$
one gets the following expression for the
bispinor wave function $ \left | \Psi \right> $ of an electron in graphene:
\begin{eqnarray}
&
\left
|\Psi\right>
=\left(
\begin{array}{c}
\widehat {\tilde\chi ^{\dagger}_{-\sigma_{_A} }}(\vec r)
\left|0,- \sigma \right>
  \\[0.6mm]
\widehat {\tilde\chi ^\dagger _{\sigma_{_B}}}(\vec r)\left|0,+\sigma\right>
\end{array}
\right)
=\left(
\begin{array}{c}
\left|0,+ \sigma\right>\widehat {\chi ^{\dagger}_{-\sigma_{_A} }}(\vec r)
\left|0,- \sigma \right>
  \\[0.6mm]
\left|0,- \sigma \right>\widehat {\chi ^\dagger _{\sigma_{_B}}}(\vec r)
\left|0,+\sigma\right>
\end{array}
\right)\nonumber
\\
&=\left(
\begin{array}{c}
\left|0,+ \sigma\right>\widehat {\chi ^{\dagger}_{-\sigma_{_A} }}(\vec r)
\left|0,- \sigma \right>
+\left|0,- \sigma \right>\widehat {\chi ^{\dagger}_{-\sigma_{_A} }}(\vec r)
\left|0,+ \sigma\right>
\\[0.6mm]
\left|0,- \sigma\right>\widehat {\chi ^\dagger _{\sigma_{_B}}}(\vec r)
\left|0,+\sigma\right> +
\left|0,+\sigma\right>\widehat {\chi ^\dagger _{\sigma_{_B}}}(\vec r)
\left|0,- \sigma\right>
\end{array}
\right)
\nonumber
\\
& =\left(
\begin{array}{c}
\widehat {\chi ^{\dagger}_{-\sigma_{_A} }} (\vec r)
\left|0,- \sigma \right> \left|0,+ \sigma \right>
 \\[0.6mm]
\widehat {\chi ^\dagger _{\sigma_{_B}}}(\vec r)
\left|0,+\sigma\right> \left|0,- \sigma\right>
\end{array}
\right)= \left(
\begin{array}{c}
\widehat {\chi ^{\dagger}_{-\sigma_{_A} }} (\vec r) \\[0.6mm]
\widehat {\chi ^\dagger _{\sigma_{_B}}}(\vec r)
\end{array}
\right)\left|0\right>
, \label{majiran-neutrino-bispinor1}
\end{eqnarray}
where
$\left|0\right> $
is a vacuum vector which consists of uncorrelated vacuum states
with spin ``down''\ $\left|0,- \sigma\right>$ and ``up''\
$\left|0,+ \sigma\right>$: $\left|0\right>=
\left|0,+\sigma\right> \left|0,- \sigma\right>$.

The density matrix $ \hat \rho _ {rr '} $ is
expressed  through the components of bispinor
(\ref{majiran-neutrino-bispinor1}) as
\begin{eqnarray}
\hat\rho _{rr'}^{AB}=\left(
\begin{array}{cc}
\widehat {\chi ^{\dagger}_{-\sigma_{_A} }} (\vec r)\, \widehat
{\chi _{\sigma_{_A} }} (\vec r) & \widehat {\chi
^{\dagger}_{-\sigma_{_A} }} (\vec r)\, \widehat {\chi
_{-\sigma_{_B}}}(\vec r\, ')
\\[1mm]
\widehat {\chi ^\dagger _{\sigma_{_B}}}(\vec r\, ')\, \widehat
{\chi _{\sigma_{_A} }} (\vec r) & \widehat {\chi ^\dagger
_{\sigma_{_B}}}(\vec r\, ')\, \widehat {\chi
_{-\sigma_{_B}}}(\vec r\, ')
\end{array}
\right) . 
\label{majiran-neutrino-density1}
\end{eqnarray}

\section{Equation for the density matrix}
In  description we will consider only  valent electrons. We
denote by $ N $ the number of atoms in two sublattices. For
valent electrons,the Dirac  hamiltonian $ H_D $ has the following
form:
\begin{eqnarray}
H_D =\sum_{L=A,B}\sum_{i=1}^{N/2}\sum_{v=1}^4\left\{
c \vec \alpha \cdot \vec p _{i_L^v} +\beta m_e c^2 -
\sum_{k=1}^N{ Ze^2\over |\vec r_{i_L^v} -
\vec R _{k}|}+
\sum_{L<L'=A,B}\sum_{i<j=1}^{N/2}\sum_{v'=1}^4
{e^2\over |\vec r_{i_L^v} -\vec r _{j_{L'}^{v'}}|}
\right\}, \nonumber 
\end{eqnarray}
\begin{eqnarray}
\label{Dirac-Hamiltonian}
\end{eqnarray}
\begin{eqnarray}
\vec p = -\imath \hbar \vec \nabla,\ \vec \alpha =\left(
\begin{array}{cc}
0& \vec \sigma \\
\vec \sigma & 0
\end{array}
\right), \  \beta =\left(
\begin{array}{cc}
1& 0 \\
0 & -1
\end{array}
\right).
\end{eqnarray}
Here $\vec \alpha, \  \beta$ is a set
$4\times 4$ of Dirac matrices,
$\vec \sigma $ is the set of
$2\times 2$ Pauli matrices,
indices
$v$ and
$v'$ enumerate
s-, p$_x$-, p$_y$
and
p$_z$ electron orbitals,
indices
$L,\ i$ and
$L',\ j$ enumerate sublattices and atoms within them respectively,
$\vec r _{i_L^v}$ is the electron radius-vector,
$\vec R _{k}$ is the radius-vector of
$k$-th carbon atom without valent electrons (atomic core),
$-Ze=-4e$ is the charge of the atomic core,
$e$ is the electron charge,
$m_e$ is the free electron mass,
$c$ is the speed of light.

The operator (matrix) of the electron density
$ \hat \rho _ {nn '; rr'} $ in the mean field approximation when
neglecting correlation interactions between electrons, satisfies the
equation
\cite{Krylova monography1,Krylova monography2}
\begin{eqnarray}
 \left(\imath {\partial \over \partial t }-(\hat h +  \Sigma^{x}+V^{sc} )\right)
 \sum_n   \hat\rho _{nn'; rr'}
=  (-\epsilon_n(0)) N_v N \delta_{rr'} \delta (t-t'),
\label{moment-Hartry-Fock-eq4}
\end{eqnarray}
where
$\hat h$ is the kinetic energy operator for a single-particle state,
$V^{sc}$ is the self-consistent potential,
$\Sigma^{x}$ is the exchange interaction,
$\epsilon_n(0)$ ($\epsilon_n(0)<0$) is the
eigenvalue of a non-excited  single-particle state
(energy of an electron orbital  for an isolated atom),
$N_v$, $N_v=4$ is the number of valent electrons.
 At
$N\to \infty $  the equation
(\ref{moment-Hartry-Fock-eq4})
can be considered as the equation for the Green function of the
quasiparticle excitations
\cite{Krylova monography1,Krylova monography2}:
\begin{eqnarray}
 \left(\imath {\partial \over \partial t }-(\hat h +  \Sigma^{x}+V^{sc} )\right)
 \sum_{n'}   \hat\rho _{nn'; rr'}
=  \delta (\vec r - \vec r\,') \delta (t-t').
\label{quasiparticleHartry-Fock-eq4}
\end{eqnarray}
Further, the  "electron"\  will be uses in the sense of a
quasiparticle.

A Dirac  -- Hartree -- Fock Hamiltonian $ H_ {DFH} $ for
quasiparticle excitations in graphene can be obtained by the
procedure of secondary quantization  of the Dirac
Hamiltonian $H_D$ (\ref{Dirac-Hamiltonian}):
\begin{eqnarray}
{1\over N_v N}
 \left(\imath {\partial \over \partial t }-\hat h_{DFH} \right)
 \sum_n   \hat\rho^{AB} _{nn'; rr'}
=  (-\epsilon_n(0))   \delta_{rr'} \delta (t-t'),
\label{Dirac-Hartry-Fock-eq}
\\
 H_{DFH}=\hat h_D +  \Sigma_{rel}^{x}+V^{sc}_{rel},
\label{Dirac-Hartry-Fock-hamiltonian}
\end{eqnarray}
where
$\hat h_D ,\ \Sigma_{rel}^{x},\ V^{sc}_{rel}$ are relativistic analogs of
operators
$\hat h ,\ \Sigma^{x},\ V^{sc}$.

The equation
(\ref{Dirac-Hartry-Fock-eq})
can be rewritten for the quasiparticle field $\chi(\vec r,  t)$ as
\begin{eqnarray}
 \left[E(p)-(\hat h_D +  \Sigma_{rel}^{x}+V_{rel}^{sc} )\right]
 \chi (\vec r,  t) =  0.
\label{quasiparticleHartry-Fock-eq5}
\end{eqnarray}
Here
$E(p)$ is the energy of the quasiparticle excitation.

Now, one can write the relativistic equation
(\ref{quasiparticleHartry-Fock-eq5}) in an explicit form:
\begin{eqnarray}
& \left(
\begin{array}{cc}
m_e c^2  - \sum_{k=1}^N{ Z e^2\over |\vec r - \vec R_k|}
                                    & c\vec \sigma \cdot \vec p\\
c\vec \sigma \cdot \vec p &
-m_e c^2 - \sum_{k=1}^N{ Z e^2\over |\vec r - \vec R_k|}
\end{array}
\right)
\left(
\begin{array}{c}
\widehat {\chi ^{\dagger}_{-\sigma_{_A} }} (\vec r) \\
\widehat {\chi ^\dagger _{\sigma_{_B}}}(\vec r)
\end{array}
\right)\left|0,-\sigma \right> \left|0,\sigma \right> 
\nonumber \\[3mm]
& + {1\over N_v\,N\, V}\sum_{i,\,i'=1}^{N_v\,N}
\int\int { d \vec r_i}{ d \vec r_{i'}} \nonumber\\[3mm]
& \times
 \left(
\begin{array}{cc}
\langle 0,-\sigma_i|
{\widehat \chi}^\dag_{-\sigma_i^A} (\vec r_i)V_{r_{i},r}
{\widehat \chi}_{\sigma_i^A}(\vec r_i)
|0,-\sigma_i\rangle
&
\langle 0,-\sigma_i|{\widehat \chi}^\dag_{-\sigma_i^A} (\vec r_i)
V_{r_{i},r}
{\widehat \chi}_{-\sigma_{i'}^B}(\vec r_i)|0,-\sigma_{i'}\rangle
\\
\langle 0,\sigma_{i'}|{\widehat \chi}^\dag_{\sigma_{i'}^B} (\vec r_{i'})
V_{r_{i'},r}
{\widehat \chi}_{\sigma_{i}^A}(\vec r_{i'})|0,\sigma_i\rangle
 &
\langle 0,\sigma_{i'}|
{\widehat \chi}^\dag_{\sigma_{i'}^B} (\vec r_{i'})
V_{r_{i'},r}
{\widehat \chi}_{-\sigma_{i'}^B}(\vec r_{i'})
|0,\sigma_{i'}\rangle
\end{array}
\right)
 \nonumber \\[3mm]
&\times
\left(
\begin{array}{c}
\widehat {\chi} ^{\dagger}_{-\sigma_{i'}{^A} } (\vec r) \\
\widehat {\chi }^\dagger _{\sigma_i{^B}}(\vec r)
\end{array}
\right)
\left|0,-\sigma \right> \left|0,\sigma \right>
=  E(p) I \left(
\begin{array}{c}
\widehat {\chi ^{\dagger}_{-\sigma_{_A} }} (\vec r) \\[0.6mm]
\widehat {\chi ^\dagger _{\sigma_{_B}}}(\vec r)
\end{array}
\right)\left|0,-\sigma \right> \left|0,\sigma \right>,
\label{quasiparticleHartry-Fock-eq6}\\[3mm]
&\widehat {\chi ^{\dagger}_{-\sigma_{_A} }} (\vec r)
={1\over N_v\,N}
\sum_{i'=1}^{N_v\,N} \widehat {\chi} ^{\dagger}_{-\sigma_{i'}{^A} } (\vec r)
,\quad
\widehat {\chi ^\dagger _{\sigma_{_B}}}(\vec r)=
{1\over N_v\,N}
\sum_{i=1}^{N_v\,N}\widehat {\chi }^\dagger _{\sigma_i{^B}}(\vec r)
\label{avarage-bispinor}
\end{eqnarray}
where 
$V_{r_{i(i')},r}\equiv V(r_{i(i')}-r)$,
$I$ is the identity matrix,
index $"i"$ ($"i'"$) enumerates all valent electrons of graphene.
From
eqs.~(\ref{quasiparticleHartry-Fock-eq6})
and
(\ref{avarage-bispinor}) the expressions follow for relativistic self-consistent Coulomb potential
$V_{rel}^{sc}$
\begin{eqnarray}
&V_{rel}^{sc}\left(
\begin{array}{c}
\widehat {\chi ^{\dagger}_{-\sigma_{_A} }} (\vec r) \\
\widehat {\chi ^\dagger _{\sigma_{_B}}}(\vec r)
\end{array}
\right)\left|0,-\sigma \right> \left|0,\sigma \right>
=
 \left(
\begin{array}{cc}
\left(V_{rel}^{sc}\right)_{AA} & 0\\
0 & \left(V_{rel}^{sc}\right)_{BB}
\end{array}
\right)
\left(
\begin{array}{c}
\widehat {\chi ^{\dagger}_{-\sigma_{_A} }} (\vec r) \\
\widehat {\chi ^\dagger _{\sigma_{_B}}}(\vec r)
\end{array}
\right)
\left|0,-\sigma \right> \left|0,\sigma \right>,
\nonumber \\
\label{self-consistent}
\\[3mm]
&\left(V_{rel}^{sc}\right)_{AA}=\sum_{i=1}^{N_v\,N}
\int { d \vec r_i}\langle 0,-\sigma_i|
{\widehat \chi}^\dag_{-\sigma_i^A} (\vec r_i)V(\vec r_i -\vec r)
{\widehat \chi}_{\sigma_i^A}(\vec r_i)
|0,-\sigma_i\rangle,
\\[3mm]
&\left(V_{rel}^{sc}\right)_{BB}=\sum_{i'=1}^{N_v\,N}\int{ d \vec r_{i'}}
\langle 0,\sigma_{i'}|
{\widehat \chi}^\dag_{\sigma_{i'}^B} (\vec r_{i'})V(\vec r_{i'} -\vec r)
{\widehat \chi}_{-\sigma_{i'}^B}(\vec r_{i'})
|0,\sigma_{i'}\rangle;
\end{eqnarray}
and exchange interaction term
$\Sigma_{rel}^{x}$ \cite{myKazan}
\begin{eqnarray}
&\Sigma_{rel}^{x}\left(
\begin{array}{c}
\widehat {\chi } ^{\dagger}_{_{-\sigma_{_A}} }(\vec r) \\
\widehat {\chi }^\dagger _{\sigma_{_B}}(\vec r)
\end{array}
\right)\left|0,-\sigma \right> \left|0,\sigma \right>
=
 \left(
\begin{array}{cc}
0& \left( \Sigma_{rel}^{x}\right)_{AB}
\\
\left( \Sigma_{rel}^{x}\right)_{BA} & 0
\end{array}
\right)
\nonumber\\
&\times\left(
\begin{array}{c}
\widehat {\chi }^{\dagger}_{-\sigma_{_A} } (\vec r) \\
\widehat {\chi} ^\dagger _{\sigma_{_B}}(\vec r)
\end{array}
\right)\left|0,-\sigma \right> \left|0,\sigma \right> \label{exchange}
, \\[3mm]
&\left( \Sigma_{rel}^{x}\right)_{AB}
\widehat {\chi }^\dagger _{\sigma_{_B}}(\vec r)\left|0,\sigma \right>
\nonumber \\
&=
\sum_{i=1}^{N_v\,N}\int { d \vec r_i}
\widehat {\chi }^\dagger _{\sigma_i{^B}}(\vec r)\left|0,\sigma \right>
\langle 0,-\sigma_i|{\widehat \chi}^\dag_{-\sigma_i^A} (\vec r_i)
V(\vec r_i -\vec r)
{\widehat \chi}_{-\sigma_B}(\vec r_i)|0,-\sigma_{i'}\rangle ,
\label{Sigma-AB}
\\[3mm]
& \left( \Sigma_{rel}^{x}\right)_{BA}
\widehat {\chi }^{\dagger}_{_{-\sigma_{_A}} } (\vec r)
\left|0,-\sigma \right>
\nonumber\\
&=\sum_{i'=1}^{N_v\,N}\int { d \vec r_{i'}}
\widehat {\chi }^{\dagger}_{_{-\sigma_{i'}^A} } (\vec r)
\left|0,-\sigma \right>
\langle 0,\sigma_{i'}|{\widehat \chi}^\dag_{\sigma_{i'}^B} (\vec r_{i'})
V(\vec r_{i'} -\vec r)
{\widehat \chi}_{_{\sigma_A}}(\vec r_{i'})|0,\sigma_i\rangle.
\label{Sigma-BA}
\end{eqnarray}

Substitution of the expressions
(\ref{self-consistent})  and
(\ref{exchange}) into eq.~
(\ref{quasiparticleHartry-Fock-eq6}) gives
\begin{eqnarray}
&\left[\left(
\begin{array}{cc}
m_e c^2  - \sum_{k=1}^N{ Z e^2 \over |\vec r - \vec R_k|}
+ \left(V_{rel}^{sc}\right)_{AA}
& c\vec \sigma \cdot \vec p + \left( \Sigma_{rel}^{x}\right)_{AB}\\[1.6mm]
c\vec \sigma \cdot \vec p + \left( \Sigma_{rel}^{x}\right)_{BA}
& -m_e c^2 -
\sum_{k=1}^N{ Z e^2\over |\vec r - \vec R_k|}
+ \left(V_{rel}^{sc}\right)_{BB}
\end{array}
 \right)  - E(p) I\right]
 \nonumber \\
& \times
\left(
\begin{array}{c}
\widehat {\chi ^{\dagger}_{-\sigma_{_A} }} (\vec r)\left|0,-\sigma \right>
\\[1.6mm]
\widehat {\chi ^\dagger _{\sigma_{_B}}}(\vec r)\left|0,\sigma \right>
\end{array}
\right)
= 0.
\label{quasiparticleDirac-Hartry-Fock-eq1}
\end{eqnarray}

Let us perform a variable change
$E\to E+m_ec^2$ and write down the system
(\ref{quasiparticleDirac-Hartry-Fock-eq1}) in components
\begin{eqnarray}
\left[ - \sum_{k=1}^N{ Z e^2\over |\vec r - \vec R_k|}
+ \left(V_{rel}^{sc}\right)_{AA}
-E(p)\right]
\widehat {\chi ^{\dagger}_{_{-\sigma_{_A} }}} (\vec r)\left|0,-\sigma \right>
\nonumber \\
+ \left[  c\vec \sigma \cdot \vec p + \left( \Sigma_{rel}^{x}\right)_{AB}\right]
\widehat {\chi ^\dagger _{_{\sigma_{_B}}}}(\vec r)\left|0,\sigma \right>
=0, \label{quasiparticleDirac-Hartry-Fock-eq2} \\
\left[ c\vec \sigma \cdot \vec p + \left( \Sigma_{rel}^{x}\right)_{BA}\right]
\widehat {\chi ^{\dagger}_{-\sigma_{_A} }} (\vec r)\left|0,-\sigma \right>
\nonumber \\
+
\left[ -2 m_e c^2 - \sum_{k=1}^N{ Z e^2\over |\vec r - \vec R_k|}
+ \left(V_{rel}^{sc}\right)_{BB} -E(p)\right]
\widehat {\chi ^\dagger _{_{\sigma_{_B}}}}(\vec r)\left|0,\sigma \right>=0.
\label{quasiparticleDirac-Hartry-Fock-eq3}
\end{eqnarray}
From the last equation of the system
(\ref{quasiparticleDirac-Hartry-Fock-eq2} --
\ref{quasiparticleDirac-Hartry-Fock-eq3}) we find the equation for the
component
$\widehat {\chi ^\dagger _{_{\sigma_{ _B}}}}(\vec r)\left|0,\sigma
\right>$
\begin{eqnarray}
\widehat {\chi ^\dagger _{_{\sigma_{_B}}}}(\vec r)\left|0,\sigma \right>
= {1\over 2 m_e c^2}
\left\{ 1 + \left[
\sum_{k=1}^N{ Z e^2\over |\vec r - \vec R_k|}
- \left(V_{rel}^{sc}\right)_{BB} +E(p)\over 2 m_e c^2
\right]
\right\}^{-1}
\nonumber\\
\times
\left[ c\vec \sigma \cdot \vec p + \left( \Sigma_{rel}^{x}\right)_{BA}\right]
\widehat {\chi ^{\dagger}_{-\sigma_{_A} }} (\vec r)\left|0,-\sigma \right>
.
\label{quasiparticleDirac-Hartry-Fock-eq4}
\end{eqnarray}


\section{
Quasirelativistic corrections}

In quasirelativistic limit
$c \to \infty $ it is possible to
 neglect lower components of bispinor on respect to upper ones,
as the components of the bispinor
$\widehat {\chi ^\dagger _{_{\sigma_{ _B}}}}(\vec r)\left|0,\sigma
\right>$
have an order of
$O(c^{-1})$.
So, it is sufficient to find  upper
components to describe the behavior of the system.

With this in mind we eliminate the small lower components in the
equation (\ref{quasiparticleDirac-Hartry-Fock-eq2}), expressing
small components through large ones  with the help of
(\ref{quasiparticleDirac-Hartry-Fock-eq4}):
\begin{eqnarray}
&\left[ - \sum_{k=1}^N{ Z e^2\over |\vec r - \vec R_k|}
+ \left(V_{rel}^{sc}\right)_{AA}
-E(p)\right]
\widehat {\chi ^{\dagger}_{-\sigma_{_A} }} (\vec r)\left|0,-\sigma \right>
+ {1\over 2 m_e c^2}
\left[  c\vec \sigma \cdot \vec p + \left( \Sigma_{rel}^{x}\right)_{AB}\right]
 \nonumber \\[1.5mm]
&\times
\left\{ 1 - \left[
-\sum_{k=1}^N{ Z e^2\over |\vec r - \vec R_k|}
+ \left(V_{rel}^{sc}\right)_{BB} -E(p)\over 2 m_e c^2
\right]
\right\}^{-1}
\left[ c\vec \sigma \cdot \vec p + \left( \Sigma_{rel}^{x}\right)_{BA}\right]
\widehat {\chi ^{\dagger}_{-\sigma_{_A} }} (\vec r)\left|0,-\sigma \right>
=0.\nonumber \\
 \label{large-bispinor-component}
\end{eqnarray}
Expanding the factor in curly brackets in a power series on a small
parameter
$$
\left|
(-{ Ze^2 / r})+ \left(V_{rel}^{sc}\right)_{BB} -E(p)\over 2 m_e c^2
\right| \ll 1 ,
$$
we obtain the quasirelativistic   Dirac -- Hartree -- Fock
approximation   for graphene:
\begin{eqnarray}
&\left[ - \sum_{k=1}^N{ Z e^2\over |\vec r - \vec R_k|}
+ \left(V_{rel}^{sc}\right)_{AA}
-E(p)\right]
\widehat {\chi ^{\dagger}_{-\sigma_{_A} }} (\vec r)\left|0,-\sigma \right>
+ {1\over 2 m_e c^2}
\left[  c\vec \sigma \cdot \vec p
+ \left( \Sigma_{rel}^{x}\right)_{AB}\right]
  \nonumber \\[3mm]
&\times
\left[ c\vec \sigma \cdot \vec p + \left( \Sigma_{rel}^{x}\right)_{BA}\right]
\widehat {\chi ^{\dagger}_{-\sigma_{_A} }} (\vec r)\left|0,-\sigma \right>
\nonumber \\[3mm]
&+ {1\over 2 m_e c^2}
\left[  c\vec \sigma \cdot \vec p + \left( \Sigma_{rel}^{x}\right)_{AB}\right]
\left[
-\sum_{k=1}^N{ Z e^2\over |\vec r - \vec R_k|}
+ \left(V_{rel}^{sc}\right)_{BB} - E(p)\over 2 m_e c^2
\right]
  \nonumber \\[3mm]
&\times
\left[ c\vec \sigma \cdot \vec p + \left( \Sigma_{rel}^{x}\right)_{BA}\right]
\widehat {\chi ^{\dagger}_{-\sigma_{_A} }} (\vec r)\left|0,-\sigma \right>
=0 \label{large-bispinor-component1} .
\end{eqnarray}

Let us find the non-relativistic limit.
With this goal in eq.~
(\ref{large-bispinor-component1}) we write down
\begin{eqnarray}
\left( \Sigma_{rel}^{x}\right)_{AB}\widehat {\chi ^{\dagger}_{-\sigma_{_A} }}
(\vec r)\left|0,-\sigma \right>
\to \left( \Sigma_{rel}^{x}\right)_{AB}
 \widehat {\chi ^{\dagger}_{_{\sigma_{_B}} }} (\vec r)
 \left|0,\sigma \right>
\label{relation_between_spinors}
\end{eqnarray}
and leave only  first order terms on  $(c^2)^{-1}$:
\begin{eqnarray}
&\left[ - \sum_{j=1}^N{ Z e^2\over |\vec r - \vec R_j|}
+ \left(V_{rel}^{sc}\right)_{AA}
-E(p)\right]
\widehat {\chi ^{\dagger}_{-\sigma_{_A} }} (\vec r)\left|0,-\sigma \right>
\nonumber \\
&+ {1\over 2 m_e c^2}
\left[  c\vec \sigma \cdot \vec p\ c\vec \sigma \cdot \vec p
+
c\vec \sigma \cdot \vec p \left( \Sigma_{rel}^{x}\right)_{BA}
+ \left( \Sigma_{rel}^{x}\right)_{AB} \left( \Sigma_{rel}^{x}\right)_{BA}
\right]
\widehat {\chi ^{\dagger}_{-\sigma_{_A} }} (\vec r)\left|0,-\sigma \right>
\nonumber \\
&+{1\over 2 m_e c^2}\left( \Sigma_{rel}^{x}\right)_{AB}c\vec \sigma \cdot \vec p
\ \widehat {\chi ^{\dagger}_{_{\sigma_{_B}} }} (\vec r)\left|0,\sigma \right>
=0 \label{nonrelativistic_limit} .
\end{eqnarray}

After some elementary algebra, we transform the equation
(\ref{nonrelativistic_limit}) to the form
\begin{eqnarray} \label{nonrelativistic_limit1}
&\left[ {\vec p\,^2\over 2 m_e }- \sum_{k=1}^N{ Z e^2\over |\vec r - \vec R_k|}
+ \left(V_{rel}^{sc}\right)_{AA}
-E(p)\right]
\widehat {\chi ^{\dagger}_{-\sigma_{_A} }} (\vec r)\left|0,-\sigma \right>
\nonumber \\[3mm]
&+ {1\over 2 }
\left[
\left( \Sigma_{rel}^{x}\right)_{BA}
\widehat {\chi ^{\dagger}_{-\sigma_{_A} }} (\vec r)\left|0,-\sigma \right>
+
\left( \Sigma_{rel}^{x}\right)_{AB}
\ \widehat {\chi ^{\dagger}_{_{\sigma_{_B}} }} (\vec r)\left|0,\sigma \right>
\right]
\nonumber \\ &
+
{1\over 2 m_e c^2}
\left( \Sigma_{rel}^{x}\right)_{AB} \left( \Sigma_{rel}^{x}\right)_{BA}
\widehat {\chi ^{\dagger}_{-\sigma_{_A} }} (\vec r)\left|0,-\sigma \right>
=0. 
\end{eqnarray}
\begin{equation}
\label{nonrelativistic_limit1}
\end{equation}

Since at replacements $ A \leftrightarrow B $ and $
\sigma_A \leftrightarrow - \sigma_B $, first two terms do not
change the form of equation, then they give the non-relativistic
contributions.
Quadratic summand
\begin{eqnarray}
\left( \Sigma_{rel}^{x}\right)_{AB} \left( \Sigma_{rel}^{x}\right)_{BA}
\label{quasirel-cjrrection}
\end{eqnarray}
is a quasirelativistic correction, because its form is sensitive to the above mentioned
change.

Since in non-relativistic limit the quasirelativistic quadratic correction
(\ref{quasirel-cjrrection})
should be omitted, the substitution of the expressions
(\ref{self-consistent}) 
and (\ref{exchange}) into eq.~(\ref{nonrelativistic_limit1}) leads to non-relativistic equation
\begin{eqnarray}
&\left[ {\vec p\,^2\over 2 m_e }- \sum_{k=1}^N{ Z e^2\over |\vec r - \vec R_k|}
+ \sum_{i=1}^{N_v\,N}
\int { d \vec r_i}\langle 0,-\sigma_i|
{\widehat \chi}^\dag_{-\sigma_i^A} (\vec r_i)
VV_{r_{i},r}
{\widehat \chi}_{\sigma_i^A}(\vec r_i)
|0,-\sigma_i\rangle
-E(p)\right]
\nonumber \\
&\times
\widehat {\chi ^{\dagger}_{-\sigma_{_A} }} (\vec r)\left|0,-\sigma \right>
+ {1\over 2 }
\left[
\sum_{i'=1}^{N_v\,N}\int { d \vec r_{i'}}
\widehat {\chi }^{\dagger}_{_{-\sigma_{i'}^A} } (\vec r) \left|0,-\sigma \right>
\langle 0,\sigma_{i'}|{\widehat \chi}^\dag_{\sigma_{i'}^B} (\vec r_{i'})
V_{r_{i'},r}
{\widehat \chi}_{_{\sigma_A}}(\vec r_{i'})|0,\sigma_i\rangle
\right.
\nonumber \\
&
\left.
+
\sum_{i=1}^{N_v\,N}\int { d \vec r_i}
\widehat {\chi }^\dagger _{\sigma_i{^B}}(\vec r)\left|0,\sigma \right>
\langle 0,-\sigma_i|{\widehat \chi}^\dag_{-\sigma_i^A} (\vec r_i)
V_{r_{i},r}
{\widehat \chi}_{-\sigma_B}(\vec r_i)|0,-\sigma_{i'}\rangle
\right]
=0 .\label{nonrelativistic_equation1}
\end{eqnarray}

Presenting $ E (p) $ as a difference
of $ m $-th energy eigenvalue for one-electron  non-excited state
$ \epsilon_m ^ {(0)} $ and the energy eigenvalue for the hole %
$\epsilon^\dag = \epsilon(p)I$: $E(p)=\epsilon_m^{(0)}-\epsilon(p)$ and
taking into account the chain of equalities
\begin{eqnarray}
\sum_{i'=1}^{N_v\,N}\int { d \vec r_{i'}}
\widehat {\chi }^{\dagger}_{_{-\sigma_{i'}^A} } (\vec r) \left|0,-\sigma \right>
\langle 0,\sigma_{i'}|{\widehat \chi}^\dag_{\sigma_{i'}^B} (\vec r_{i'})
V(\vec r_{i'} -\vec r)
{\widehat \chi}_{_{\sigma_A}}(\vec r_{i'})|0,\sigma_i\rangle
\nonumber \\
\equiv
\sum_{i'=1}^{N_v\,N}\int { d \vec r_{i'}}
\widehat {\chi }^{\dagger}_{_{-\sigma_{i'}^A} } (\vec r) \left|0,-\sigma \right>
\langle 0,\sigma_{i'}|{\widehat \chi}^\dag_{\sigma_{i'}^B} (\vec r_{i'})
V(\vec r_{i'} -\vec r)
{\widehat \chi}_{_{-\sigma_B}}(\vec r_{i'})|0,-\sigma_{i'}\rangle
\nonumber \\
=
\sum_{i'=1}^{N_v\,N}\int { d \vec r_{i'}}
\widehat {\chi }^{\dagger}_{_{\sigma_{i'}^B} } (\vec r) \left|0,-\sigma \right>
\langle 0,-\sigma_{i'}|{\widehat \chi}^\dag_{-\sigma_{i'}^A} (\vec r_{i'})
V(\vec r_{i'} -\vec r)
{\widehat \chi}_{_{-\sigma_B}}(\vec r_{i'})|0,-\sigma_{i'}\rangle ,
\end{eqnarray}
one can rewrite the equation
(\ref{nonrelativistic_equation1}) as
\begin{eqnarray}
&\left[ {\vec p\,^2\over 2 m_e }-
\sum_{k=1}^N{ Z e^2\over |\vec r - \vec R_k|}
\right.\nonumber \\
&\left.
+ \sum_{i=1}^{N_v\,N}
\int { d \vec r_i}\langle 0,-\sigma_i|
{\widehat \chi}^\dag_{-\sigma_i^A} (\vec r_i)V(\vec r_i -\vec r)
{\widehat \chi}_{\sigma_i^A}(\vec r_i)
|0,-\sigma_i\rangle
-\left(\epsilon_m^{(0)}-\epsilon(p)\right)\right]
\nonumber \\
&\times
\widehat {\chi ^{\dagger}_{-\sigma_{_A} }} (\vec r)\left|0,-\sigma \right>
+
\sum_{i=1}^{N_v\,N}\int { d \vec r_i}\nonumber \\
& \times
\widehat {\chi }^\dagger _{\sigma_i{^B}}(\vec r)\left|0,-\sigma \right>
\langle 0,-\sigma_i|{\widehat \chi}^\dag_{-\sigma_i^A} (\vec r_i)
V(\vec r_i -\vec r)
{\widehat \chi}_{-\sigma_B}(\vec r_i)|0,-\sigma_{i}\rangle
=0 \label{nonrelativistic_equation2} .
\end{eqnarray}
As mentioned above, in non-relativistic  limit the indices $ A, \ B $ can
be omitted and  eq.~(\ref{nonrelativistic_equation2}) can be written  in a
final form
\begin{eqnarray}
& \epsilon_m^{(0)}
\widehat \psi^\dag_{\sigma_m}(\vec r_m) |0,\sigma_m\rangle
- \langle 0,-\sigma_i|
\widehat \epsilon^\dag\ \widehat { \mbox{I}}|0,-\sigma_i\rangle
\widehat \psi^\dag_{\sigma_m} (\vec r_m)
|0,\sigma_m\rangle  %
  \nonumber\\
&=
 \hat h (\vec r_m) {\widehat \psi }_{\sigma_m}(\vec r_m)|0,\sigma_m\rangle
  \nonumber\\
&  -
\sum_{i=1}^n \int { d \vec r_i}
{\widehat \psi}^\dag _{\sigma_i}(\vec r_m)|0,\sigma_m\rangle
V(\vec r_i -\vec r_m)
\langle 0,-\sigma_i|{\widehat \psi}^\dag_{-\sigma_i} (\vec r_i)
{\widehat \psi}_{\sigma_m}(\vec r_i)|0,-\sigma_i\rangle
  \nonumber\\
&\hspace{-10mm} + \sum_{i=1}^n \int { d \vec r_i}
{\widehat \psi}^\dag _{\sigma_m}(\vec r_m) |0,\sigma_m\rangle
V(\vec r_i -\vec r_m)\langle 0,-\sigma_i|
{\widehat \psi}^\dag_{-\sigma_i} (\vec r_i)
{\widehat \psi}_{\sigma_i}(\vec r_i)
|0,-\sigma_i\rangle
    \      
 \label{Geisenberg-motion-eq7}
\end{eqnarray}
where
$\sigma_m\equiv -\sigma$, $\vec r_m \equiv  \vec r$,
\begin{eqnarray}
\hat h (\vec r_m)=
{\vec p\,^2\over 2 m_e }- \sum_{k=1}^N{ Z e^2\over |\vec r_m - \vec R_k|} .
\end{eqnarray}
The formula
(\ref{Geisenberg-motion-eq7}) represents
precisely 
the
Hartree -- Fock equation for the  spin
electron density as it was shown in
\cite{myKazan}.

Thus, distinction of spinor wave functions of the electrons belonging
to different sublattices, are manifested through the interaction of the
sublattices.
The spin dependence of Dirac cones is manifested in the first order in
$ (c ^ 2) ^ {-1} $ when one can not neglect the lower components of the
bispinor.
When neglecting the small lower spinor components,
the description becomes non-relativistic, and therefore does not allow to
describe
spin-dependent polarization of the  band structure of graphene and
graphene-like material.

Now, it is possible to consider the energy-band structure of graphene with the
second quantized Hamiltonian
(\ref{Dirac-Hartry-Fock-hamiltonian}).
We choose the Bloch functions
\begin{equation}
\chi_n(\vec k, \vec r)= e^{\imath \vec k \cdot \vec r} u_n( \vec r)
\label{Bloch-function}
\end{equation}
as a basis ones to describe a wave function of
quasiparticles in graphene. The
Bloch function with a wave vector $ \vec k $ at point with radius-vector
$ \vec r $ has the form
\begin{eqnarray}
\chi_n(\vec k, \vec r)= {1\over (2\pi)^{3/2}\sqrt{2N}}
\sum_{\vec R_{l}} e^{i \vec k \cdot \vec R_{l}}
\psi_{\{n\}}(\vec r -\vec R_{l}),
\end{eqnarray}
where
$\psi_{\{n\}}$ is the atomic orbital with a set of quantum numbers
 $\{n\}$.

A wave function
$\Psi(\vec k, \vec r)$ of an electron in graphene has the form
\begin{eqnarray}
\Psi(\vec k, \vec r)=
 {1\over \sqrt{ 2}}\sum_n
\left[
 c_A^n\chi_n^{(A)}(\vec k, \vec r)+
c_B^n\chi_n^{(B)}(\vec k, \vec r)
\right]
\end{eqnarray}
with the normmalization condition given by
{\Large ${ \int }$} ${\left|\Psi(\vec k, \vec
r)\right|^2 d \vec r }= \sum_n\left[(c_A^n)^2+
(c_B^n)^2\right]=1$.

As a zero order approximation
$ \psi_m ^ {(0)} (\kappa r) $, $\kappa r \equiv \vec \kappa \cdot \vec r $
for functions $ \psi_ {\{m \}} $ we adopt the solution of a single-electron
problem for an isolated atom.
As the number of electrons for   $ C $  atom is even, there are no
pseudo-potential terms in the self-consistent Hartree -- Fock equation \cite{Fock}:
\begin{eqnarray}
\left[ h(\vec r)
 +  \hat V^{sc} (\kappa r) - \hat \Sigma ^x(\kappa r) \right]
 \psi_m^{(0)} (\kappa r)
 =
 \epsilon_m^{(0)}\psi_m^{(0)} (\kappa r),
 \label{hartry-fock-eqs1}
\end{eqnarray}
where
$\epsilon_m^{(0)}$ is the $m$-th eigenvalue of the Hamilton operator for a
single-electron state of the isolated  atom C.

\section{Brillouin zone corner approximation}
Let us consider peculiar points in momentum space for graphene.
These points correspond to the diffraction peaks, the so-called
reflexes of the diffraction pattern.
These are the
corners $K$  and $K'$ of the graphene
Brillouin hexagonal zone, which we designate as
$K_A$  and $K_B$, respectively. Their positions are given by \cite{Neto}
\begin{equation}
\begin{split}
\vec K_A =\left( {2\pi\over 3a},\ {2\pi\over 3\sqrt{3} a},\ 0\right),
\
\vec K_B =\left( {2\pi\over 3a},\ -{2\pi\over 3\sqrt{3} a},\ 0\right).
\end{split}
\end{equation}
Here $a \approx 1.44$ \AA \ is the carbon-carbon distance.

The basis Bloch function
$\chi_n^{(i)}(\vec k^{(i)}, \vec r)$, $i=A,\ B$
for the description of an electron in one of the sublattices has the form
\begin{eqnarray}
\chi_n^{(i)}(\vec k^{(i)}, \vec r)= {1\over (2\pi)^{3/2}\sqrt{N/2}}
\sum_{\vec R^{(i)}_{l}} \exp\{i \vec k^{(i)} \cdot \vec R^{(i)}_{l}\}
\psi_{\{n\}}(\vec r -\vec R^{(i)}_{l}). \label{approx-wave-function}
\end{eqnarray}

Let us make a variables change
\begin{eqnarray}
\vec k^{(i)} = \vec K_i-(\vec K_i- \vec k^{(i)})\equiv
\vec K_i - \vec q
.
\label{corner-change}
\end{eqnarray}
Taking into account of the change (\ref{corner-change
})
the wave functions (\ref{approx-wave-function}) in the
primitive subcell of the graphene space are approximately described as
\begin{eqnarray}
\chi_n^{(i)}(\vec K_i - \vec q, \vec r)= {1\over (2\pi)^{3/2}\sqrt{N/2}}
\sum_{\vec R^{(i)}_{l}}
\exp\{i [\vec K_i - \vec q\, ] \cdot \vec R^{(i)}_{l}\}
\psi_{\{n\}}(\vec r -\vec R^{(i)}_{l}). \label{approx-wave-function1}
\end{eqnarray}

\section{Secondary quantized Hamiltonian of  quasi-two-dimensional graphene\label{sect}}

Now we construct the secondary quantized Hamiltonian $ H_ {qu} $ in this
approximation.
For this purpose, we rewrite the expression (\ref{approx-wave-function1})
in the form (\ref {Bloch-function}):
\begin{eqnarray}
&\chi_n^{(i)}(\vec K_i - \vec q, \vec r)=
\exp\{i [\vec K_i - \vec q\, ] \cdot \vec r\}
{1\over (2\pi)^{3/2}\sqrt{N/2}}
\sum_{\vec R^{(i)}_{l}}
\exp\{i [\vec K_i - \vec q\, ] \cdot [\vec R^{(i)}_{l}-\vec r]\}
\nonumber \\
&\times \psi_{\{n\}}(\vec r -\vec R^{(i)}_{l})
\equiv \exp\{i [\vec K_i - \vec q\, ] \cdot \vec r\}
\Psi_n^{(i)}( \vec r)
. \label{approx-wave-function2}
\end{eqnarray}

Left multiplying eq.~(\ref{nonrelativistic_limit})
on the Dirac bra-vector
$\left<0,\sigma \right|\widehat {\chi _{_{-\sigma_{_B}} }} (\vec r)$,
we find that
\begin{eqnarray}
&\left<0,\sigma \right|\widehat {\chi _{_{-\sigma_{_B}} }} (\vec r)
\left[ - \sum_{j=1}^N{ Z e^2\over |\vec r - \vec R_j|}
+ \left(V_{rel}^{sc}\right)_{AA}
-E(p)\right]
\widehat {\chi ^{\dagger}_{-\sigma_{_A} }} (\vec r)\left|0,-\sigma \right>
\nonumber \\
&+ \left<0,\sigma \right|\widehat {\chi _{_{-\sigma_{_B}} }} (\vec r)
{1\over 2 m_e c^2}
\left[  c\vec \sigma \cdot \vec p\ c\vec \sigma \cdot \vec p
+
c\vec \sigma \cdot \vec p \left( \Sigma_{rel}^{x}\right)_{BA}
+ \left( \Sigma_{rel}^{x}\right)_{AB} \left( \Sigma_{rel}^{x}\right)_{BA}
\right]
\nonumber \\
&
\times
\widehat {\chi ^{\dagger}_{-\sigma_{_A} }} (\vec r)\left|0,-\sigma \right>
+\left<0,\sigma \right|\widehat {\chi _{_{-\sigma_{_B}} }} (\vec r)
{1\over 2 m_e c^2}\left( \Sigma_{rel}^{x}
\right)_{AB}c\vec \sigma \cdot \vec p
\
\widehat {\chi ^{\dagger}_{_{\sigma_{_B}} }} (\vec r)\left|0,\sigma \right>
=0 \label{massless-Dirac-eq} .
\end{eqnarray}
Let us consider quasi-two-dimensional model of graphene, when the
radius-vectors $ \vec r $ deviate slightly from the plane of the
monolayer. Therefore, values of $ q $ are small: $ q <1 $, and one can
omit terms of order $q^2$. In this
case, the use of  (\ref{approx-wave-function2}) allows to transform the multiplier
$$
\left[\left<0,\sigma \right|\widehat {\chi _{_{-\sigma_{_B}} }} (\vec r)
\vec \sigma \cdot \vec \nabla\ \right]
\left[
\vec \sigma \cdot \vec \nabla
\widehat {\chi ^{\dagger}_{-\sigma_{_A} }} (\vec r)\left|0,-\sigma \right>
\right]
$$
in
eq.~(\ref{massless-Dirac-eq}) to the following form:
\begin{eqnarray}
-\vec \sigma \cdot(\vec q  - \vec K_B)\vec \sigma \cdot
(\vec K_A - \vec q\,)
\approx
(\vec \sigma \cdot \vec K_A )( \vec \sigma \cdot \vec K_B)
-  (\vec \sigma \cdot \vec q \,)(\vec \sigma \cdot
(\vec K_A + \vec K_B)).
\label{massless-Dirac-factor}
\end{eqnarray}

Let we renormalize the energy as follows:
\begin{eqnarray}
E \to \tilde E -{ \hbar^2\over 2 m_e}
(\vec \sigma \cdot \vec K_A  )(\vec \sigma \cdot \vec K_B).
\label{energy-renormalization}
\end{eqnarray}
Substitution of
(\ref{massless-Dirac-factor}, \ref{energy-renormalization})
into eq.~(\ref{massless-Dirac-eq}) gives the following equation:
\begin{eqnarray}
&\left<0,\sigma \right|\widehat {\chi _{_{-\sigma_{_B}} }} (\vec r)
\left[ - \sum_{j=1}^N{ Z e^2\over |\vec r - \vec R_j|}
+ \left(V_{rel}^{sc}\right)_{AA}
\right]
\widehat {\chi ^{\dagger}_{-\sigma_{_A} }} (\vec r)\left|0,-\sigma \right>
\nonumber \\
&+ \left<0,\sigma \right|\widehat {\chi _{_{-\sigma_{_B}} }} (\vec r)
{1\over 2 m_e c^2}
\left[  c^2 \hbar ^2 (\vec \sigma \cdot \vec q \,)(\vec \sigma \cdot
(\vec K_A + \vec K_B)
\right. \nonumber \\
&\left. +
c\vec \sigma \cdot \vec p \left( \Sigma_{rel}^{x}\right)_{BA}
+ \left( \Sigma_{rel}^{x}\right)_{AB} \left( \Sigma_{rel}^{x}\right)_{BA}
\right]
\widehat {\chi ^{\dagger}_{-\sigma_{_A} }} (\vec r)\left|0,-\sigma \right>
+{1\over 2 m_e c^2}
\left<0,\sigma \right|\widehat {\chi _{_{-\sigma_{_B}} }} (\vec r)
\nonumber \\
&\times
\left( \Sigma_{rel}^{x}
\right)_{AB}c\vec \sigma \cdot \vec p
\
\widehat {\chi ^{\dagger}_{_{\sigma_{_B}} }} (\vec r)\left|0,\sigma \right>
=
\left<0,\sigma \right|\widehat {\chi _{_{-\sigma_{_B}} }} (\vec r)
\tilde E(p)
\widehat {\chi ^{\dagger}_{-\sigma_{_A} }} (\vec r)\left|0,-\sigma \right>
\label{massless-Dirac-eq1} .
\end{eqnarray}
Since
\begin{eqnarray}
\hbar  (\vec \sigma \cdot \vec q \,)
\widehat {\chi ^{\dagger}_{-\sigma_{_A} }} (\vec r)\left|0,-\sigma \right>
=
- i \hbar  (\vec \sigma \cdot \vec \nabla )
\widehat {\chi ^{\dagger}_{-\sigma_{_A} }} (\vec r)\left|0,-\sigma \right>
,
\end{eqnarray}
 the equation
(\ref{massless-Dirac-eq1}) can be transformed to the form
\begin{eqnarray}
&\left<0,\sigma \right|\widehat {\chi _{_{-\sigma_{_B}} }} (\vec r)
\left[ - \sum_{j=1}^N{ Z e^2\over |\vec r - \vec R_j|}
+ \left(V_{rel}^{sc}\right)_{AA}
\right]
\widehat {\chi ^{\dagger}_{-\sigma_{_A} }} (\vec r)\left|0,-\sigma \right>
\nonumber \\
& + \left<0,\sigma \right|\widehat {\chi _{_{-\sigma_{_B}} }} (\vec r)
{1\over 2 m_e c^2}
\left[  c^2 \hbar (- i \hbar  (\vec \sigma \cdot \vec \nabla ))
 (\vec \sigma \cdot
(\vec K_A + \vec K_B)
\right. \nonumber \\
&\left. +
c\vec \sigma \cdot \vec p \left( \Sigma_{rel}^{x}\right)_{BA}
+ \left( \Sigma_{rel}^{x}\right)_{AB} \left( \Sigma_{rel}^{x}\right)_{BA}
\right]
\widehat {\chi ^{\dagger}_{-\sigma_{_A} }} (\vec r)\left|0,-\sigma \right>
+{1\over 2 m_e c^2}
\left<0,\sigma \right|\widehat {\chi _{_{-\sigma_{_B}} }} (\vec r)
\nonumber \\
&\times \left( \Sigma_{rel}^{x}
\right)_{AB}c\vec \sigma \cdot \vec p
\
\widehat {\chi ^{\dagger}_{_{\sigma_{_B}} }} (\vec r)\left|0,\sigma \right>
=
\left<0,\sigma \right|\widehat {\chi _{_{-\sigma_{_B}} }} (\vec r)
\tilde E(p)
\widehat {\chi ^{\dagger}_{-\sigma_{_A} }} (\vec r)\left|0,-\sigma \right>
\label{massless-Dirac-eq2} .
\end{eqnarray}

The overlap integrals for the same sublattices are much smaller than that
for different sublattices.
Furthermore, for the quasi-two-dimensional graphene, the first term in
the left-hand
side of eq.~(\ref{massless-Dirac-eq2}) describing the screening
is also small.
Therefore, we can neglect the first and the last terms in the left-hand
side of eq.~(\ref{massless-Dirac-eq2}):
\begin{eqnarray}
& \left<0,\sigma \right|\widehat {\chi _{_{-\sigma_{_B}} }} (\vec r)
{1\over 2 m_e c^2}
\left\{  c\vec \sigma \cdot \vec p
\left[
\left( \Sigma_{rel}^{x}\right)_{BA}+ c \hbar \vec \sigma \cdot
(\vec K_A + \vec K_B)\right]
\right.\nonumber\\
&\left.
+ \left( \Sigma_{rel}^{x}\right)_{AB} \left( \Sigma_{rel}^{x}\right)_{BA}
\right\}
\widehat {\chi ^{\dagger}_{-\sigma_{_A} }} (\vec r)\left|0,-\sigma \right>
=
\left<0,\sigma \right|\widehat {\chi _{_{-\sigma_{_B}} }} (\vec r)
\tilde E(p)
\widehat {\chi ^{\dagger}_{-\sigma_{_A} }} (\vec r)\left|0,-\sigma \right>
\label{massless-Dirac-eq3} .
\end{eqnarray}
The left-hand side of eq.~(\ref{massless-Dirac-eq3}) represents
itself  matrix elements of the secondary quantized Hamiltonian
$ H_{qu} $ of quasi-two-dimensional graphene, in which the motion
of quasiparticle excitations of electronic subsystem is described
by the equation of the form
\begin{eqnarray}
& \left\{  \vec \sigma \cdot \vec p
\left[
\left( \Sigma_{rel}^{x}\right)_{BA}+ c \hbar \vec \sigma \cdot
(\vec K_A + \vec K_B)\right]
+ {1\over c } \left( \Sigma_{rel}^{x}\right)_{AB} \left( \Sigma_{rel}^{x}\right)_{BA}
\right\}
\widehat {\chi ^{\dagger}_{-\sigma_{_A} }} (\vec r)\left|0,-\sigma \right>
\nonumber \\
&=
E_{qu}(p)
\widehat {\chi ^{\dagger}_{-\sigma_{_A} }} (\vec r)\left|0,-\sigma \right>
\label{massless-Dirac-eq4} ,
\end{eqnarray}
and
$E_{qu}(p)$ is defined by the expression
\begin{equation}
E_{qu}(p) = 2 m_e c \tilde E(p).
\end{equation}
The equation
(\ref{massless-Dirac-eq4}) is nothing but an equation which describes
the motion of Dirac charge carrier in the quasi-two-dimensional graphene:
\begin{eqnarray}
 \left\{   \vec \sigma \cdot \vec p \ \hat v^{qu}_F
- {1\over c } \left( i\Sigma_{rel}^{x}\right)_{AB} \left( i\Sigma_{rel}^{x}\right)_{BA}
\right\}
\widehat {\chi ^{\dagger}_{-\sigma_{_A} }} (\vec r)\left|0,-\sigma \right>
=
E_{qu}(p)
\widehat {\chi ^{\dagger}_{-\sigma_{_A} }} (\vec r)\left|0,-\sigma \right>
\label{massless-Dirac-eq5}
\end{eqnarray}
where operator
$  \hat v^{qu}_F$ is defined as
\begin{equation}
  \hat v^{qu}_F =  \left[
\left( \Sigma_{rel}^{x}\right)_{BA}+ c \hbar \vec \sigma \cdot
(\vec K_A + \vec K_B)\right]. \label{Fermi-velocity}
\end{equation}
The physical meaning of the quasirelativistic corrections
(\ref{quasirel-cjrrection}), entered in (\ref{massless-Dirac-eq5}),
is in appearance of a pseudo-mass $ m_ {pseudo} $ for the charge
carriers in graphene:
\begin{eqnarray}
m_{-}={1 \over c}
\left( i\Sigma_{rel}^{x}\right)_{AB} \left( i\Sigma_{rel}^{x}\right)_{BA}.
\label{quasirel-cjrrection1}
\end{eqnarray}

Since quasirelativistic  correction (\ref{quasirel-cjrrection}) is included with a small factor
$ c ^ {-1} $, then it may be neglected and one obtains the equation of motion for a massless quasiparticle
charge carrier in quasi-two-dimensional graphene:
\begin{eqnarray}
 \left\{  \vec \sigma \cdot    \vec p \ \hat v^{qu}_F
 \right\}
\widehat {\chi ^{\dagger}_{-\sigma_{_A} }} (\vec r)\left|0,-\sigma \right>
=
E_{qu}(p)
\widehat {\chi ^{\dagger}_{-\sigma_{_A} }} (\vec r)\left|0,-\sigma \right>.
\label{massless-Dirac-eq6}
\end{eqnarray}
According to (\ref{massless-Dirac-eq6}), the massless charge carrier moves
with the Fermi velocity operator $\hat v^{qu} _F$ (\ref{Fermi-velocity}).

Transforming the Fermi velocity operator
$\hat v^{qu}_F$ (\ref{Fermi-velocity})
to a matrix form one arrives to different values of Fermi velocity in
different directions.

\section{Charge carriers asymmetry}

The operator of pseudo-mass
(\ref{quasirel-cjrrection1}) is not invariant in respect to
transformation
$A \to B $:
\begin{eqnarray}
m_{- }\
\widehat {\chi ^{\dagger}_{-\sigma_{_A} }} (\vec r)\left|0,-\sigma \right>
\neq {1 \over c}
\left(i \Sigma_{rel}^{x}\right)_{BA} \left(i \Sigma_{rel}^{x}\right)_{AB}
\widehat {\chi ^{\dagger}_{\sigma_{_B} }} (\vec r)
\left|0,\sigma \right>
\stackrel{def}{=\!=} m_{+}\
\widehat {\chi ^{\dagger}_{\sigma_{_B} }} (\vec r)\left|0,\sigma \right>
\label{pseudomass}
\end{eqnarray}
where
$m_{+  }$ is a hole mass in graphene.
Due to the factor
$c^{-1}$, pseudomass
$m_{\mp}$ in
(\ref{pseudomass}) is small.

Energy can be calculated based on the following equation:
\begin{eqnarray}
E_{qu}(p) =\left<0,\sigma \right|\widehat {\chi _{_{-\sigma_{_B}} }} (\vec r)
    \vec \sigma \cdot \vec p \ \hat v^{qu}_F
\widehat {\chi ^{\dagger}_{-\sigma_{_A} }} (\vec r)
\left|0,-\sigma \right> \nonumber \\
+{1\over c }
\left<0,\sigma \right|\widehat {\chi _{_{-\sigma_{_B}} }}(\vec r)
 \left( \Sigma_{rel}^{x}\right)_{AB} \left( \Sigma_{rel}^{x}\right)_{BA}
\widehat {\chi ^{\dagger}_{-\sigma_{_A} }} (\vec r)
\left|0,-\sigma \right>
\label{energy-dispersion} .
\end{eqnarray}
From the last, one arrives to the energy dispersion law for graphene:
\begin{eqnarray}
E_{qu}^2(p) =\left<0,\sigma \right|\widehat {\chi _{_{-\sigma_{_B}} }} (\vec r)
    p \ \hat v^{qu}_F
\widehat {\chi ^{\dagger}_{-\sigma_{_A} }} (\vec r)
\left|0,-\sigma \right>^2 \nonumber \\
+{1\over c^2 }
\left<0,\sigma \right|\widehat {\chi _{_{-\sigma_{_B}} }}(\vec r)
 \left( \Sigma_{rel}^{x}\right)_{AB} \left( \Sigma_{rel}^{x}\right)_{BA}
\widehat {\chi ^{\dagger}_{-\sigma_{_A} }} (\vec r)
\left|0,-\sigma \right>^2
\label{energy-dispersion} .
\end{eqnarray}

Let us represent the dispersion law
(\ref{energy-dispersion}) in dimensionless form
\begin{eqnarray}
{E_{-}\over m_ev_F^2}
={1\over v_F^2} \sqrt{
{v^2_F p^2\over m_e^2}+
\left<0,\sigma \right|\widehat {\chi _{_{-\sigma_{_B}} }}(\vec r)
 \left(\hat v_F^{qu}\right)^4
\widehat {\chi ^{\dagger}_{-\sigma_{_A} }} (\vec r)
\left|0,-\sigma \right>^2
}.
\label{energy-dispersion1}
\end{eqnarray}
Similar expression can be written for holes.

Performing series expansion of
(\ref{energy-dispersion1}) one arrives at
\begin{eqnarray}
{E_{\mp}\over m_ev_F^2}
\approx {\hbar\over v_F^2}\left[
{v_F k\over m_e}\pm {1\over 2}{m_e
\left<0,\sigma \right|\widehat {\chi _{_{-\sigma_{_B}} }}(\vec r)
 \left(\hat v_F^{qu}\right)^4
 \widehat {\chi ^{\dagger}_{-\sigma_{_A} }} (\vec r)
 \left|0,-\sigma \right>^2 \over \hbar^2 v_F k
}\right].
\label{energy-dispersion2}
\end{eqnarray}

Next, utilizing  the expression for the wave function
(\ref{approx-wave-function1}) we estimate
matrix element
$$
\left<0,\sigma \right|\widehat {\chi _{_{-\sigma_{_B}} }}(\vec r)
  \left(\hat v_F^{qu} \right)^4 \widehat {\chi ^{\dagger}_{-\sigma_{_A} }} (\vec r)
 \left|0,-\sigma \right>
$$
in tight-binding approximation:
\begin{eqnarray}
\left<0,\sigma \right|\widehat {\chi _{_{-\sigma_{_B}} }}(\vec r)
  \left(\hat v_F^{qu} \right)^4
   \widehat {\chi ^{\dagger}_{-\sigma_{_A} }} (\vec r)
 \left|0,-\sigma \right>=
{1\over (2\pi)^{3/2}\sqrt{N/2}} \int d^2r
 \left(\chi_n^{B}\right)^\dagger \nonumber\\
 \times   \hat v_F^{qu}
\sum_{\vec R^{A}_{l}}
\exp\{i [\vec K_{A} - \vec k\, ] \cdot \vec R^{A}_{l}\}
\psi_{\{n\}}(\vec r -\vec R^{A}_{l})\approx
{N[V_{WS}/2]^{-1}1\over (2\pi)^{3/2}\sqrt{N/2}}
\int_{V_{WS}/2} d^2r
\nonumber\\
\times \int_{V_{BZ}} d^2k'
 \left(\chi_n^{B}\right)^\dagger
  \left(\hat v_F^{qu} \right)^4 \sum_{\vec \delta_{l}}
\exp\{i [\vec K_{A} - \vec k + \vec k'\, ] \cdot \vec \delta_{l}\}
\psi_{\{n\}}(\vec {\tilde k } -\vec K_A ) e^{i(\vec {\tilde k} -\vec K_A)\cdot \vec r}
\label{matrix-element}
\end{eqnarray}
where $\vec \delta_1 = {a\over 2} (1,\ \sqrt{3})$,
$\vec \delta_2 = {a\over 2} (1,\ \sqrt{3})$, and
$\vec \delta_3 = -a (1,\ 0)$
are three nearest-neighbor vectors in real lattice space whose length
$\delta_i$, $i=1,\ 2, 3$ is
given by
\begin{equation}
\delta_1= \delta_2 =\delta_3 =a , \label{length}
\end{equation}
$V_{WS}$ is a volume of the Wigner--Seitz 2D-cell, $V_{BZ}$ is a volume of
the Brillouin 2D-zone, $\vec {\tilde k } -\vec K_A =  \vec k'$.

Using the facts that
$\sum_{\vec \delta_i} e^{i\vec K_{A}\cdot\vec \delta_i}=0$ and
$$
e^{i(\vec K_{A}-\vec k + \vec k'\,)\cdot\vec \delta_i} \simeq
 (1+ik'\delta_i)[e^{i\vec K_{A}\cdot\vec \delta_i} - i \sin (\vec k\cdot \vec \delta_i)
 \cos (\vec K_A \cdot \vec \delta_i)]
\simeq  i\vec k'\cdot \vec\delta_i e^{i\vec K_{A}\cdot\vec \delta_i}- i a k
+ a^2 kk'$$
for $k\ll 1$ and $a\to 0 $,
performing inverse Fourier transformation
for the function
$\psi_{\{n\}}(\vec {\tilde k } -\vec K_A )$
and integrating over $d^2r$ one derives that the following expression
takes place in a vicinity of the Dirac point:
\begin{eqnarray}
\left<0,\sigma \right|\widehat {\chi _{_{-\sigma_{_B}} }}(\vec r)
  (\hat v_F^{qu} )^4\widehat {\chi ^{\dagger}_{-\sigma_{_A} }} (\vec r)
 \left|0,-\sigma \right>\approx
{N[V_{WS}/2]^{-1}\over (2\pi)^{3/2}\sqrt{N/2}}
\delta (\vec {\tilde k } -\vec K_A) \nonumber \\
\times \int d^2r'\,
 \left(\chi_n^{B}\right)^\dagger
  \left(\hat v_F^{qu} \right)^4 \sum_{l=1}^3
\int_{-\vec {\tilde k}}^{\vec {\tilde k}}
[i k'\delta_l  (\cos\theta_{\vec k\, ',\ \vec \delta_l})
-  a k ]\ d\vec {k'}\
\psi_{\{n\}}(\vec r\, ')
 . \label{matrix-element0-1}
\end{eqnarray}

Now, let us expand operators
$\hat \chi$, $\hat \chi^\dagger $ entering into $\left(
\Sigma_{rel}^{x}\right)_{BA}$ (\ref{Sigma-BA}), $\left(
\Sigma_{rel}^{x}\right)_{AB}$ (\ref{Sigma-AB}), respectively, up
to  linear order in $\vec \delta_i$. We act by the operator
expansions
$\left( \Sigma_{rel}^{x}\right)_{B(A)A(B)}$,
which enter
in $(\hat v_F^{qu})^4$,
on appropriate wave functions
$\psi_{\{n\}}$, and
use an explicit form of the wave function
$\chi_n^{B}$, analogous
to (\ref{approx-wave-function1}). As a result, we get
\begin{eqnarray}
&\left<0,\sigma \right|\widehat {\chi _{_{-\sigma_{_B}} }}(\vec r)
  (\hat v_F^{qu} )^4\widehat {\chi ^{\dagger}_{-\sigma_{_A} }} (\vec r)
 \left|0,-\sigma \right>\approx
{\delta_{ \vec {\tilde k },\ \vec K_A}
\over V_{WS}}
\sum_{\vec R^{A}_{j},\ \vec R^{B}_{j}}\exp\{- i\vec k\,  \cdot
(\vec R^{A}_{j}- \vec R^{B}_{j})\}
  \nonumber \\
&\times
\left[\int d^2r\, '\,
 |\psi_{n}|^2(\vec r\, ')\right] [V_{WS}/2]^4
  \left[ \int_{V_{WS}/2} \psi^\dagger_{n}(\vec r - \delta _i )
V(r) \psi_{n}(\vec r )d\vec r
\right]^4 \nonumber \\
&\times
\sum_{l,\ n,\ m,\ p =1}^3 {1\over (2\pi)^3}
\int_{-\vec {\tilde k}}^{\vec {\tilde k}}
 [k'^5\delta_l ^4 (\cos\theta_{\vec k',\ \vec \delta_l})^4
- {k'}^4 \delta_l ^3 (\cos\theta_{\vec k',\ \vec \delta_l})^3 a k ]\ dk'\ .
 \label{matrix-element1}
\end{eqnarray}
Since the integral of odd function
$\int_{-\vec {\tilde k}}^{\vec {\tilde k}}
 [k'^5\delta_l ^4 (\cos\theta_{\vec k',\ \vec \delta_l})^4 ]\ dk'$
entering into (\ref{matrix-element1})
vanishes
and
$\int \, |\psi_{n}|^2(\vec r\, ')d^2r\, '= 1 $, we find the desired
matrix element
\begin{eqnarray}
\left<0,\sigma \right|\widehat {\chi _{_{-\sigma_{_B}} }}(\vec r)
  (\hat v_F^{qu} )^4\widehat {\chi ^{\dagger}_{-\sigma_{_A} }} (\vec r)
 \left|0,-\sigma \right>\approx
{\delta_{ \vec {\tilde k },\ \vec K_A}}  
 [3   V_{WS}/2]^4 a k { (\tilde k)^5}
\nonumber\\
\times
  \left[ \int_{V_{WS}/2} \psi^\dagger_{n}(\vec r - \delta _i )
V(r) \psi_{n}(\vec r )d\vec r
\right]^4 
 {3a^3\over V_{W}}
 \approx - 3  (v_F)^4 ak
 \label{matrix-element2}
\end{eqnarray}
where $v_F={3a\over 2}\int_{V_{WS}/2} \psi^\dagger_{n}(\vec r - \delta _i )
V(r) \psi_{n}(\vec r )d\vec r$, $V_{WS}\approx a^2$.

Substituting
(\ref{matrix-element2} ) into (\ref{energy-dispersion2})
we obtain the energy dispersion law in graphene
\begin{eqnarray}
{E_{\mp}\over m_ev_F^2} = {\hbar\over m_e }
\left({k\over v_F} \mp {1\over 2}3 a {m_e^2 v_F\over  \hbar^2}
\right).
 \label{energy-dispersion3}
\end{eqnarray}

The discrepancy between the energy dispersion laws of  electrons and
holes
due to their different pseudo-masses will lead to discrepancy, though
small,
of  cyclotron mass $m^*_\pm$ of holes and electrons.

Within the semiclassical approximation 
\cite{Ashcroft} 
the cyclotron mass is defined by the formula
\begin{eqnarray}
{m^* \over m_e} = {1\over 2\pi m_e} \left. \left[
{\partial S(E)\over \partial E}\right]\right|_{E=E_F}=
{E_F\over m_e v_F^2}, \label{cyclotron-mass}
\end{eqnarray}
where the area $S(E)$ in momentum space enclosed by the orbit  is given
as $S(E)=\pi q(E)^2=\pi E^2/v_F^2$, $v_F$ is ordinary scalar Fermi
velocity in graphene:
$v_F\simeq  10^6 $~m/s.
Addition of a small mass term leads to electron-hole asymmetry
of the dependence of  cyclotron mass upon  charge carrier concentration.

Let us estimate this asymmetry based on the concentration dependence of cyclotron mass
on respect to the free-electron mass $ m_e $.

With this  in mind,
let us substitute
(\ref{energy-dispersion3}) into
(\ref{cyclotron-mass})
and find the difference of cyclotron masses of charge carriers
\begin{eqnarray}
\Delta {m^*}_{theor}/m_e \sim  3 {\hbar\over m_e }
 a {m_e^2 v_F\over  \hbar^2}\approx 3 {1.44\hbar\over m_e }
 \label{asymmetry}
\end{eqnarray}
where $\Delta {m^*}=m^*_+ - m^*_- $

Cyclotron mass $m^*_\pm$ of charge carriers in graphene  has
been extracted from the temperature dependence of the
Shubnikov--de Haas oscillations
in 
\cite{Novoselov}.
The experimental dependencies
of cyclotron masses $m_+^*$, $m_-^*$ of holes
and electron upon concentration
$n_+^{exp}$ and $n_-^{exp}$ of the charged carriers %
have been fitted based on a dimensionless formula
for massless quasiparticles:
\begin{eqnarray}
{m^*_\pm\over m_e}= {E_F^\pm\over v^2_F m_e}=
{\hbar k_F^\pm\over v_F m_e}= {\hbar \sqrt{\pi }\over v_F m_e}
\sqrt{n^{exp}_\pm}
 \label{experimental-asymmetry}
\end{eqnarray}
when accounting
the relation
between the surface density $n^{exp}_\pm $ and the Fermi momentum $k_F^\pm $ in the form
$(k_F^\pm) ^2 /\pi = n^{exp}_\pm $.
Taking into account of small difference of the pseudo-masses
$m_\pm$ an estimation of the experimental data based
on the formula (\ref{experimental-asymmetry})
gives the  value $\Delta {m^*}_{exp}$ of the asymmetry
in a form
\begin{eqnarray}
\Delta {m^*}_{exp}/m_e  = {\hbar \sqrt{\pi }\over v_F m_e}
(\sqrt{n^{exp}_+} - \sqrt{n^{exp}_-})
= {\hbar \sqrt{\pi }\over v_F m_e}
(\sqrt{n^{exp}_- +\Delta n} - \sqrt{n^{exp}_-})\nonumber
\\
\simeq {\hbar \sqrt{\pi }\over v_F m_e}
{\Delta n\over 2 \sqrt{n^{exp}_-}}
\simeq
3 {\hbar \sqrt{\pi }\over   m_e}
 \label{experimental-asymmetry1}
\end{eqnarray}
where $\Delta n $ is the experimental value.

Practical coincidence  of experimental (\ref{experimental-asymmetry1}) and
theoretical
(\ref{asymmetry}) estimates of electron-hole asymmetry demonstrates friutfullness
of the developed approach and the valuable argument to support
the assumption that charge carriers in graphene are pseudo-Dirac quasiparticles having
small but finite pseudo-mass.

\section{Dirac cone  and replicas in graphene}
Today for the description of scattering in graphene the
pseudo-massless Dirac approximation is used
\cite{Katsnelson2006b,Katsnelson2006c,Katsnelson2006-et-al}.
In this approximation, the Dirac cone is double-degenerated.

The non-relativistic replicas  in the Hartree -- Fock
self-consistent field approximation are degenerated also as in the
massless pseudo-relativistic approximation.

As it was established above, quasirelativistic correction perturbs the
system
and removes the degeneracy of  atomic wave functions for C$_A $, C$_B $.
Besides, due to the hexagonal lattice symmetry one of these Dirac cones is
a replica repeated six times.
The replicas form a hexagonal mini  Brillouin zone around the primary
Dirac cone.

In what follows we estimate the effects of Dirac cone splitting on
primary Dirac cone and its replicas. There will be also considered the influence
of operator of the Fermi velocity $\hat v^{qu}_F$ introduced in section \ref{sect} on the form of dispersion law in graphene.

\section{Two-dimensional graphene}
Consider a model of graphene, when the radius-vectors $ \vec r $ practically do not
  leave  the plane of the monolayer.
Such a model is applicable  for very small
$\vec q $, $q \ll 1 $.
Therefore, for two-dimensional graphene in the vicinity of corners  $ K_i $
one can set $ z $-component of vectors equal to zero $\vec r =(x,\ y,\ 0)$,
$\vec q =(q_x,\ q_y,\ 0)$, $q \ll 1 $.
Since the vectors lie in one plane, near the corners
$ K_i $ it is possible
  to transform the expression (\ref{massless-Dirac-factor}) to the following
form
\begin{eqnarray}
 (\vec \sigma \cdot \vec q \,)(\vec \sigma \cdot
(\vec K_A + \vec K_B)) %
= \vec q \cdot (\vec K_A + \vec K_B) +
i \vec \sigma \cdot[(\vec K_A + \vec K_B)\times \vec q\, ]
\nonumber \\
\simeq  \vec q \cdot (\vec K_A + \vec K_B) +
    i \sigma_z [(\vec K_A + \vec K_B)\times \vec q\, ]_z, \quad q\ll 1.
 \label{KK-mixing}
\end{eqnarray}
Accounting of
(\ref{KK-mixing}) and neglecting the quasi-relativistic
correction
(\ref{quasirel-cjrrection}), it is possible to simplify the equation
(\ref{massless-Dirac-eq6}) in the following way:
\begin{eqnarray}
 \left\{ \vec \sigma _{2D} \cdot \vec p \, \hat v_F  +
\vec q \cdot (\vec K_A + \vec K_B) + i \sigma_z [(\vec K_A + \vec K_B)
\times \vec q\, ]_z
 \right\}
\widehat {\chi ^{\dagger}_{-\sigma_{_A} }} (\vec r)\left|0,-\sigma \right>
\nonumber \\
= E_{qu}(p)
\widehat {\chi ^{\dagger}_{-\sigma_{_A} }} (\vec r)\left|0,-\sigma \right>
, \ q\ll 1
\label{massless-Dirac-eq7}
\end{eqnarray}
where
$\vec \sigma_{2D}$ is the two-dimensional
vector of the Pauli matrices:
$\vec \sigma_{2D} = (\sigma_x, \ \sigma_y) $,
operator of two-dimensional Fermi velocity
$\hat v_F$ is defined by the expression
\begin{equation}
\hat v_F =   \left( \Sigma_{rel}^{x}\right)_{BA}.
\end{equation}

Let us demonstrate that the presence of operator
$\hat v_F$ rather than scalar
$v_F$,
leads to a rotation of Dirac cone on respect to replicas.
To do this,
we perform the following non-unitary transformation of the wave function
for graphene
\begin{equation}
\widehat {\tilde \chi ^{\dagger}_{-\sigma_{_A} }}
\left|0,-\sigma \right> =
\left( \Sigma_{rel}^{x}\right)_{BA}
\widehat {\chi ^{\dagger}_{-\sigma_{_A} }} \left|0,-\sigma \right>.
\end{equation}
After this transformation, eq.~
(\ref{massless-Dirac-eq7})
takes the form similar to  pseudo-Dirac approximation of two-dimensional
graphene:
\begin{eqnarray}
 \left\{ \vec \sigma _{2D}^{AB} \cdot \vec p_{BA}   +
\vec q \cdot (\vec K_A + \vec K_B) + i \sigma_z^{BA} [(\vec K_A + \vec K_B)\times \vec q\, ]_z
 \right\}
\widehat {\tilde
\chi ^{\dagger}_{-\sigma_{_A} }} (\vec r)\left|0,-\sigma \right>
\nonumber \\
=
E_{qu}(p)
\widehat {\tilde \chi ^{\dagger}_{-\sigma_{_A} }} (\vec r)
\left|0,-\sigma \right>
, \ q\ll 1
\label{massless-Dirac-eq8}
\end{eqnarray}
where
$\vec \sigma _{2D}^{AB}= \left( \Sigma_{rel}^{x}\right)_{BA}
\vec \sigma _{2D}
\left( \Sigma_{rel}^{x}\right)_{BA}^{-1}$,
$\vec p_{BA}\widehat {\tilde \chi ^{\dagger}_{-\sigma_{_A} }}=
\left( \Sigma_{rel}^{x}\right)_{BA} \vec p\,
\left( \Sigma_{rel}^{x}\right)_{BA}^{-1}
\widehat {\tilde \chi ^{\dagger}_{-\sigma_{_A} }}$.
Due to the fact that
$\left( \Sigma_{rel}^{x}\right)_{BA}\neq
\left( \Sigma_{rel}^{x}\right)_{AB}$,
the vector $\vec p_{BA}$ of the Dirac cone axis
 is somehow rotated in respect to the vector $\vec p_{AB}$ of its replica
as qualitatively shown in Fig.~
2. We investigate this in detail a bit later.

The term with a scalar product
$\vec q \cdot (\vec K_A + \vec K_B)$ in eq.~
(\ref{KK-mixing})  shifts equally the bands of quasi-particles,
whereas the term with the vector product leads to different
sign of shift for bands of electrons and holes due to the presence of
$\sigma_z$.
  The pseudo-mass $m_-$ (\ref{quasirel-cjrrection1}) with corrections on
these shifts leads to appearance of an energy gap $\Delta _{BA}$
between valent and conduction Dirac zones (Fig.~
2).
By analogy, the pseudo-mass $m_+$ (\ref{pseudomass}) with corrections on
these shifts leads to
appearance of an energy gap $\Delta _{AB}$, $\Delta _{AB} > \Delta_{BA}$
between valent and conduction  zones of the Dirac replica (see Fig.~
2).
\begin{figure}[hbt]\label{fig2}
\includegraphics[width=8.cm,height=6.cm,angle=0]{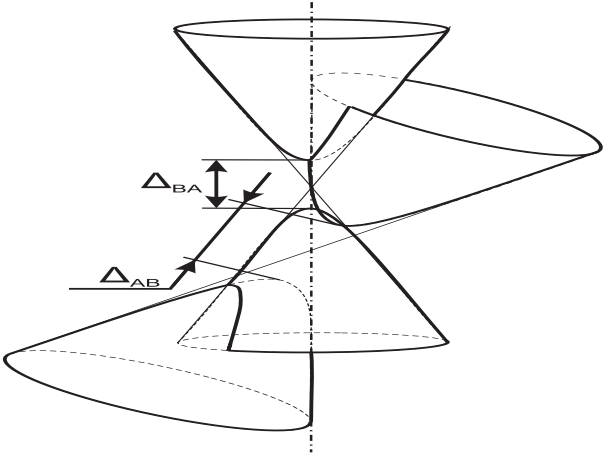}
\caption{Qualitative representation of splitting for double-degenerated Dirac cone. }
\end{figure}
As one can see from Fig.~
2, the conduction  zone of the replica leads to
the gap  between valent and conduction  zones of the primary
Dirac cone. Therefore,
in spite of the gap, the monolayer graphene is semimetal.
This theoretically predicted effect can explains qualitatively the
high-intensity narrow strip connecting the valence and the conduction band
of
graphene in the vicinity of the Dirac point, which has been experimentally
observed in
 ARPES spectra of the monolayer epitaxial graphene on
SiC(0001) 
\cite{Zhou,
Gweon}.

Besides, due to diagonality of the Pauli matrix $\sigma_z$,
a mixing of waves
$\Psi_{K_A}$ and $\Psi_{K_B}$ takes place.

In real experiments investigating the electron motion
in the vicinity of the corners
$K_A, \ K_B$ of the graphene Brillouin zone,
the value of $q$
is not an infinitely small but gets some finite though small values.
Therefore, the effect of solutions mixing in the vicinity of points
$K_A, \ K_B$ is always presented though it could be small enough.
According the results
(\ref{Fermi-velocity},
\ref{massless-Dirac-eq6}) of the previous section,
the effect of the mixing is a manifestation of the graphene lattice
anisotropy.
To existing physical systems whose properties are strongly depended
on the presence
of graphene lattice anisotropy
one can refers to graphene nanoribbons with zigzag
and armchair edges, including carbon nanotubes
\cite{Brey,
Nakada,
maksimenko}. 
It is stipulated by the fact that, due to the finite width,  these systems
effectively represent
quasi-one-dimensional configurational ones
\cite{Krylova monography1,Krylova monography2}.
The maximal mixing effect should be expected for charge carriers motion in
graphene nanoribbons with zigzag and armchair
edges.

Thus, the proposed graphene model can qualitatively explain
experimentally observed
different electrical conductivity of
zigzag and armchair  graphene nanoribbons as a result of electron-holes asymmetry of graphene.

Now let us estimate the effects stipuleter by the Fermi velocity operator $\hat v_F$.

\section{Approximation of $\pi$ (p$_z$) electrons}
We consider electrons in monolayer graphene in commanly used assumption on the absence
of mixing of states for Dirac points
$K_A$ and $K_B$.
Then the eq.~(\ref{massless-Dirac-eq8}) and eq.~
\begin{eqnarray}\label{eigen}
 \vec \sigma _{2D}^{BA} \cdot \vec p_{AB}
\widehat {\tilde
\chi ^{\dagger}_{+\sigma_{_B} }} (\vec r)\left|0,-\sigma \right>
=
E_{qu}(p)
\widehat {\tilde \chi ^{\dagger}_{+\sigma_{_B} }} (\vec r)
\left|0,-\sigma \right>
, \ q\ll 1
\label{pseudi-Dirac-whithout-mix2}
\end{eqnarray}
describe the delocalized
$\pi$-electron on a Dirac cone and its replicas.
Here
$\vec \sigma _{2D}^{BA}= \left( \Sigma_{rel}^{x}\right)_{AB}
\vec \sigma _{2D}
\left( \Sigma_{rel}^{x}\right)_{AB}^{-1}$,
$\vec p_{AB}\widehat {\tilde \chi ^{\dagger}_{+\sigma_{_B} }}=
\left( \Sigma_{rel}^{x}\right)_{AB} \vec p\,
\left( \Sigma_{rel}^{x}\right)_{AB}^{-1}
\widehat {\tilde \chi ^{\dagger}_{+\sigma_{_B} }}$,

\noindent $\widehat {\tilde \chi ^{\dagger}_{+\sigma_{_B} }} (\vec r)
\left|0,\sigma \right>=
\left( \Sigma_{rel}^{x}\right)_{AB}
\widehat {\chi ^{\dagger}_{+\sigma_{_B} }} (\vec r)
\left|0,\sigma \right>$.

We write down  wave functions $\phi_{\uparrow}$ and $\phi_{\downarrow}$ with spin ``up'' and ``donwn'' appropriately
in the form
\begin{eqnarray}
\phi_{\uparrow} ={1\over \sqrt{2}}\left(
\begin{array}{c}
\phi_1\\
0
\end{array}
\right),\qquad
\phi_{\downarrow} ={1\over \sqrt{2}}\left(
\begin{array}{c}
0\\
\phi_2
\end{array}
\right).
\end{eqnarray}

To obtain numerical estimate but without full scale {\it ab initio} simulations
we restrict ourselves by consideration of non-selfconsistence problem.
With this in mind,
the bispinor wave functions of quasi-particles (in the vicinity of Dirac cone)
are presented as (almost) free massless Dirac field (p$_{z}$-electrons):
\begin{eqnarray}
&\left(
\begin{array}{c}
\widehat {\chi ^{\dagger}_{-\sigma_{_{A(B)}} }} (\vec r)
\left|0,-\sigma \right>\\
\widehat {\chi ^{\dagger}_{\sigma_{_{B(A)}} }} (\vec r)
\left|0,\sigma \right>
\end{array}
\right) ={e^{-\imath (\vec K_{A(B)} -\vec q)\cdot \vec r}\over \sqrt{2}}\left(
\begin{array}{c}
\exp\{- \imath \theta_{k_{A(B)}}\}\phi_1\\
\exp\{-\imath \theta_{k_{A(B)}}\}\phi_2\\
-\exp\{\imath \theta_{k_{B(A)}}\}\phi_2\\
\exp\{\imath \theta_{k_{B(A)}}\}\phi_1
\end{array}
\right)\nonumber\\
&\equiv \left(
\begin{array}{c}
\chi_{-\sigma_{A(B)}}\\
\chi_{+\sigma_{B(A)}}
\end{array}
\right),\label{pi-electronA_down} \\
&\left(
\begin{array}{c}
\widehat {\chi ^{\dagger}_{+\sigma_{_{A(B)}} }} (\vec r)
\left|0,\sigma \right>\\
\widehat {\chi ^{\dagger}_{-\sigma_{_{B(A)}} }} (\vec r)
\left|0,-\sigma \right>
\end{array}
\right) ={e^{-\imath (\vec K_{A(B)} -\vec q)\cdot \vec r}\over \sqrt{2}}
\left(
\begin{array}{c}
-\exp\{\imath \theta_{k_{A(B)}}\}\phi_2\\
\exp\{\imath \theta_{k_{A(B)}}\}\phi_1\\
\exp\{- \imath \theta_{k_{B(A)}}\}\phi_1\\
\exp\{- \imath \theta_{k_{B(A)}}\}\phi_2
\end{array}
\right)\nonumber\\
&
\equiv \left(
\begin{array}{c}
\chi_{+\sigma_{A(B)}}\\
\chi_{-\sigma_{B(A)}}
\end{array}
\right) \label{pi-electronA_up}
\end{eqnarray}
where
\begin{equation}
\phi_i=
{1\over (2\pi)^{3/2}\sqrt{N/2}}
\sum_{\vec R^{A(B)}_{l}}
\exp\{\imath [\vec K_{A(B)} - \vec q\, ] \cdot [\vec R^{A(B)}_{l}-\vec r]\}
\psi_{\{n_i\}}(\vec r -\vec R^{A(B)}_{l})
. \label{Bloch-function-reduc}
\end{equation}
This form is coherent with known scattering problem considerations \cite{Katsnelson2006-et-al} when all $\phi_i$ should be placed as unity.

Then, in accord with formulas (\ref{pi-electronA_down},
\ref{pi-electronA_up})
the expression for the wave function (\ref{avarage-bispinor}) is transformed into the following
\begin{eqnarray}
\widehat {\chi} ^{\dagger}_{-\sigma_{i'}{^A} } (\vec r)
\equiv \widehat {\chi ^{\dagger}_{-\sigma_{_A} }} (\vec r)
,\quad
\widehat {\chi }^\dagger _{\sigma_i{^B}}(\vec r) \equiv
\widehat {\chi ^\dagger _{\sigma_{_B}}}(\vec r) \mbox{ for }
\forall \ i,\ i'.
\label{avarage-bispinor1}
\end{eqnarray}

Now, it is possible to write down matrices
$\left( \Sigma_{rel}^{x}\right)_{AB}\approx \Sigma_{AB}$ and
$ \left( \Sigma_{rel}^{x}\right)_{BA} \approx \Sigma_{BA}$
without sef-action, e.g. for $\Sigma_{AB}$ as:
\begin{eqnarray}
&\left( \Sigma_{rel}^{x}\right)_{AB}
\widehat {\chi ^{\dagger}_{+\sigma_{_B} }} (\vec r)
\left|0,\sigma \right> \approx   \Sigma_{AB}\ \chi_{\sigma_B}
=
\sum_{i=1}^{N_v N-1} \int d\vec r_i
\nonumber \\
&\times  V(\vec r_i -\vec r)
\left[
\chi_{-\sigma_A} (\vec r_i)\cdot
\chi_{-\sigma_B}^*( \vec r_i)
\right]\chi_{+\sigma_B} (\vec r)
=
{1\over 2^{3/2}} \sum_{i=1}^{N_v N-1} \int d\vec r_i V(\vec r_i -\vec r)
 \nonumber \\
& \times
\left[
\begin{array}{cc}
e^{-\imath  \theta_{k_A}} \phi_1 (\vec r_i)
e^{\imath  \theta_{k_B}} \phi^*_1(\vec r_i)
&
e^{-\imath  \theta_{k_A}} \phi_1 (\vec r_i)
e^{\imath  \theta_{k_B}} \phi^*_2 (\vec r_i)\\
e^{-\imath  \theta_{k_A}} \phi_2 (\vec r_i)
e^{\imath  \theta_{k_B}} \phi^*_1 (\vec r_i)&
e^{-\imath  \theta_{k_A}}\phi_2(\vec r_i)
e^{\imath  \theta_{k_B}} \phi^*_2 (\vec r_i)
\end{array}
\right]
\left[
\begin{array}{c}
- e^{-\imath [(\vec K_A -\vec q)\cdot \vec r - \theta_{k_B}]}
\phi_2 (\vec r) \\
e^{-\imath [(\vec K_A -\vec q)\cdot \vec r - \theta_{k_B}]}
\phi_1(\vec r)
\end{array} \nonumber
\right],\\
\end{eqnarray}
and similar expression for $\Sigma_{BA}$.

Now we use the tight-bidning approximation for further simplification of
 $\Sigma_{AB}$ and $\Sigma_{BA}$ matrices.
Accounting only nearest neighbors and choosing
$\psi_{\{n_2\}},\ \psi_{\{n_1\}}$ as orbitals of $\pi$-electron:
$$\psi_{\{n_2\}}=c_1\psi_{\mbox{\small p}_z}(\vec r\pm \vec \delta_i)+
c_2\psi_{\mbox{\small p}_z}(\vec r), \sum_{i=1}^2 c_i=1;\
\psi_{\{n_1\}}= \psi_{\mbox{\small p}_z}(\vec r),$$
after some algebra we gets
\begin{eqnarray}
&\Sigma_{AB} ={1\over \sqrt{2}(2\pi)^{3}}
e^{-\imath (\theta_{k_A}-\theta_{k_B})}
\sum_{i=1}^{3}
\exp\{\imath [\vec K_A - \vec q\, ] \cdot \vec \delta_i\}
\int  V(\vec r) d\vec r
\nonumber \\
&\times \left(
\begin{array}{cc}
\sqrt{2}\psi_{\mbox{\small p}_z} (\vec r )
\psi^*_{\mbox{\small p}_z, - \vec \delta_i} (\vec r )
 &
\psi_{\mbox{\small p}_z} (\vec r )
[\psi^*_{\mbox{\small p}_z}(\vec r)
+ \psi^*_{\mbox{\small p}_z, - \vec \delta_i}(\vec r)]\\
\psi^*_{\mbox{\small p}_z, - \vec \delta_i} (\vec r )
[\psi_{\mbox{\small p}_z, \vec \delta_i}(\vec r)+
\psi_{\mbox{\small p}_z}(\vec r)]
&
{[\psi_{\mbox{\small p}_z, \vec \delta_i}(\vec r)+
\psi_{\mbox{\small p}_z}(\vec r)]
[\psi^*_{\mbox{\small p}_z}(\vec r)
+\psi^*_{\mbox{\small p}_z, - \vec \delta_i}(\vec r)]
\over \sqrt{2}}
\end{array}
\right), \nonumber
\end{eqnarray}
\begin{eqnarray}
\label{Sigma-AB3}
\end{eqnarray}
\begin{eqnarray}
&  \Sigma_{BA} = {1\over \sqrt{2}(2\pi)^{3}} e^{-\imath
(\theta_{k_A}-\theta_{k_B})}
\sum_{i=1}^{3}
\exp\{\imath [\vec K_A - \vec q\, ] \cdot \vec \delta_i\}
\int  V({\vec r}) d {\vec r}
\nonumber \\
&\times \left(
\begin{array}{cc}
{[\psi_{\mbox{\small p}_z, \vec \delta_i}(\vec r)+
\psi_{\mbox{\small p}_z}(\vec r)]
[\psi^*_{\mbox{\small p}_z}(\vec r)
+\psi^*_{\mbox{\small p}_z, - \vec \delta_i}(\vec r)]
\over \sqrt{2}}
 &
-
\psi^*_{\mbox{\small p}_z, - \vec \delta_i} (\vec r )
[\psi_{\mbox{\small p}_z, \vec \delta_i}(\vec r)+
\psi_{\mbox{\small p}_z}(\vec r)]
\\
-
\psi_{\mbox{\small p}_z} (\vec r)
[\psi^*_{\mbox{\small p}_z}(\vec r)
+ \psi^*_{\mbox{\small p}_z, - \vec \delta_i}(\vec r)] &
\sqrt{2}\psi_{\mbox{\small p}_z} (\vec r )
\psi^*_{\mbox{\small p}_z, - \vec \delta_i} (\vec r)
\end{array}
\right).\nonumber
\end{eqnarray}
\begin{eqnarray}
 \label{Sigma-BA3}
\end{eqnarray}
Here $c_1=c_2=1/\sqrt{2}$, we choose the upper sign for $\pi$-orbital
$\psi_{\{n_2\}}$ and introduce a notion
$\psi_{\mbox{\small p}_z,\ \pm \vec \delta_i}(\vec r_{2D})
=\psi_{\mbox{\small p}_z}(\vec r_{2D}\pm \vec \delta_i)$.

Substituting known expression for eigenfunctions of hydrogen-like atom, evaluating integrals
we obtain rather lengthly ($\vec q$)-dependent invertible matrices, for $q=0$ they are pure numeric and up to a common scalar prefactor
read
\begin{eqnarray}\label{sab0}
  \Sigma_{AB}(\vec K_A)=\left(
\begin{array}{cc}
 -0.88 & -4.3 \\
 -1.4 & -4.1
\end{array}
\right)
\end{eqnarray}

\begin{eqnarray}\label{sba0}
  \Sigma_{BA}(\vec K_A)=\left(
\begin{array}{cc}
 -4.1 & 1.4 \\
 4.3 & -0.88
\end{array}
\right).
\end{eqnarray}

\begin{figure}[hbt]\label{spectrum}
\hspace{3cm} (a) \hspace{7cm} (b) \\
\includegraphics[width=8.cm,angle=0]{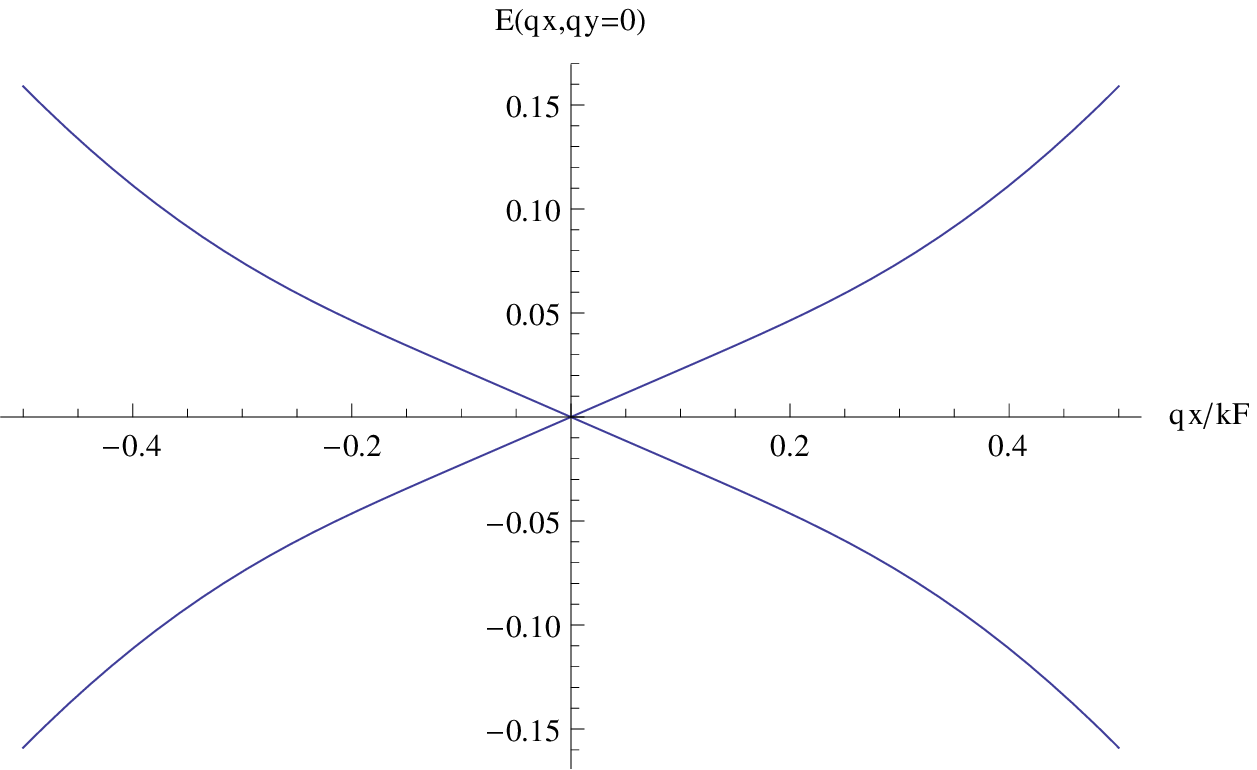}\hspace{0.2cm}
\includegraphics[width=8.cm,angle=0]{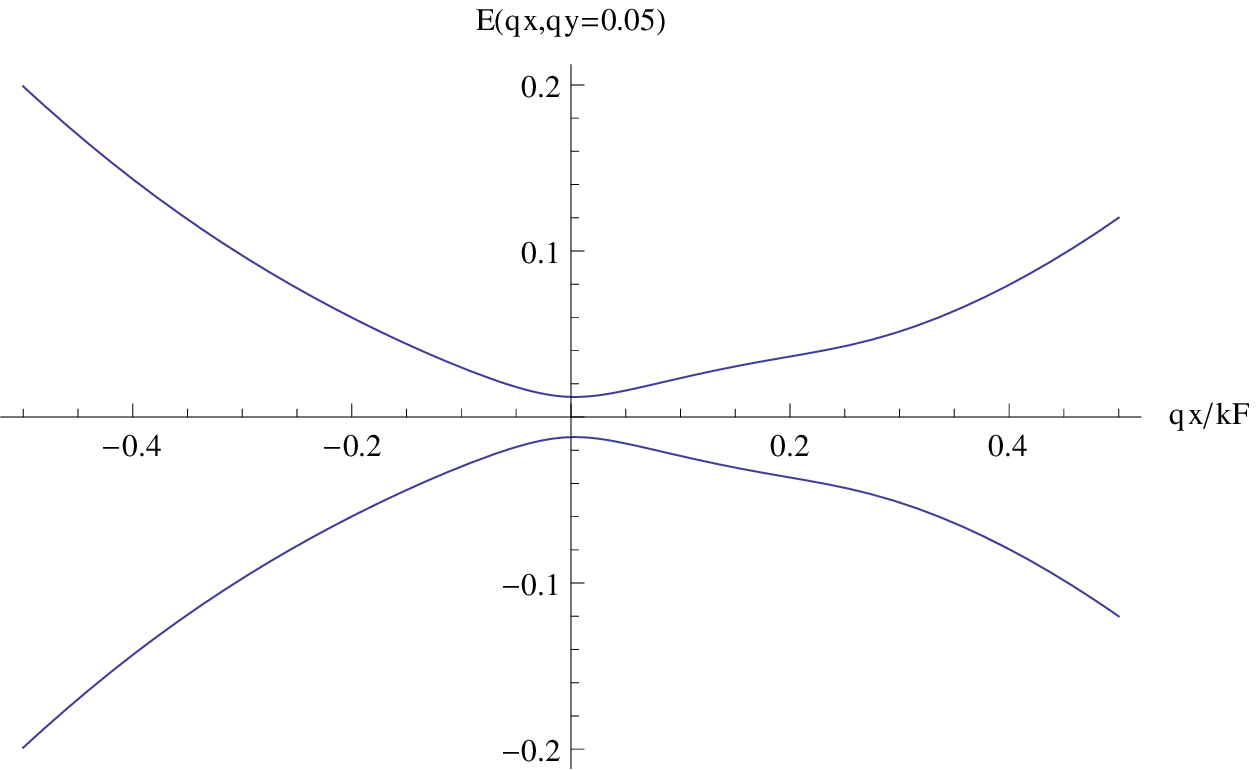}\hspace{0.2cm}
\\ \ \\
\ \hspace{8cm} (c) \ \\
\includegraphics[width=8.cm,angle=0]{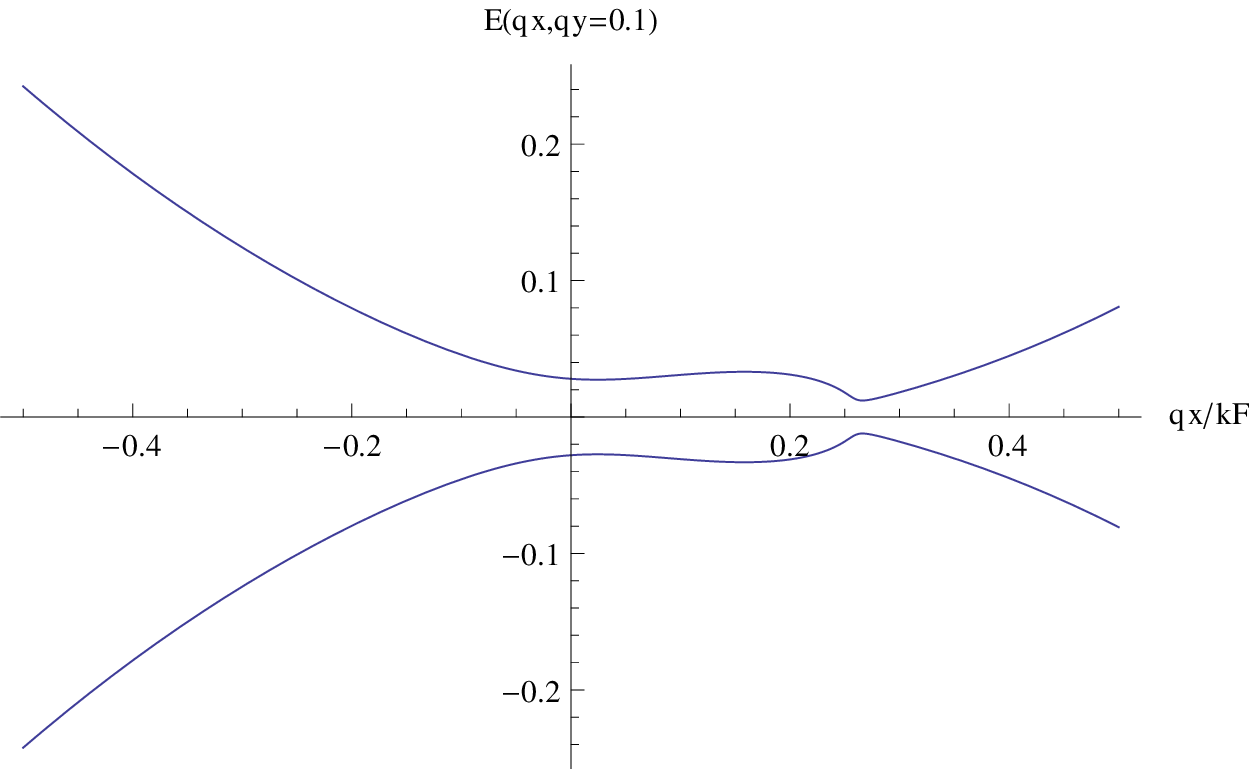}
\caption{The dependence $E(q_x,q_y)$ for several different values
of $q_y$: (a) -- $q_y/k_F=0$, (b) -- $q_y/k_F=0.05$, (c) --
$q_y/k_F=0.1$
 }
\end{figure}

Taking into account the validity condition for our approximation $|\vec q|\ll 1$, performing series expansion on $\vec q$
up to a linear terms, we get e.g. for $\Sigma_{AB}$.
\begin{eqnarray}\label{sigmaab}
&  \Sigma_{AB}\nonumber \\
&= \left(
\begin{array}{cc}
 (0.43 +0.33 i){q_x}+(0.25 -0.57 i) {q_y}-0.88 & (2.1+1.6 i){q_x}+(1.2 -2.8 i){q_y}-4.3 \\
 (0.74 +0.73 i){q_x}+(0.43 -1.3 i){q_y}-1.4    & (2.0+1.7 i){q_x}+(1.2 -2.9 i){q_y}-4.1
\end{array}
\right). %
\nonumber \\
\end{eqnarray}
The most interesting thing in (\ref{sigmaab}) is that eigenvalue problem (\ref{eigen}) gives precisely the known dispersion laws
$E(\vec q)=\pm \sqrt{q_x^2+q_y^2}$, that is problem is persistent for linear in $|\vec q|$ variations.

Now, we take into account higher order in $|\vec q|$ terms when
evaluating $\Sigma_{AB},\ \Sigma_{BA}$. The spectrum corresponding
to eq.~(\ref{eigen}) deviates from the conic form, that we
demonstrate by $E(\vec q)$ surface sections for few $q_y$ in
Fig.~
3.
 Fig.~
3a emphasizes  that in the vicinity of Dirac point the cone is
persistent due to symmetry (section crosses original cone and its
replicas simultaneously),  at higher values of $q$, higher order
corrections start to contribute. When section crosses original
cone only (Fig.~
3b) we find higher order corrections to charge
carrier asymmetry. In Fig.~
3c we can observe the dispersion curve corresponded to the section
crossing  original Dirac cone and one of its replicas. We can
estimate the $q$-distance between the $K_A$ point and one of its
reflexes in ARPES experiments as a distance from the origin to
$q_x$ value corresponded to the local curve minimum near
$q_x/k_F=0.27$ (and minigap $E_{mg}\approx 0.02\ k_F v_F$), but of
course, approximations we made were too rough to obtain a credible
result, it should be considered as qualitative only.

As it has been shown above divergence of Dirac cone and its replicas relative to each
other leads to the electron-hole asymmetry.
Then, the six fold rotational symmetry of graphene near the Dirac
point energy breaks.
The displacement of replicas points in the graphene
Brillouin zone on respect to the primary Dirac cone points occurs on a
distance
\begin{equation}
|\Delta \vec q_{AB} | = |\vec q_{AB} - \vec q_{BA}|.
\label{rotate}
\end{equation}
Since in the neighborhood of the top of the Dirac cone $q \to 0 $,
and (as was shown above) the cone persists then, in accord with
(\ref{rotate}) in the graphene Brillouin zone  all points of the
Dirac cone replicas shift, except of their tops (see Fig.~
4).

\begin{figure}[hbt]\label{fig3}
\includegraphics[width=8.4cm,height=6.cm,angle=0]{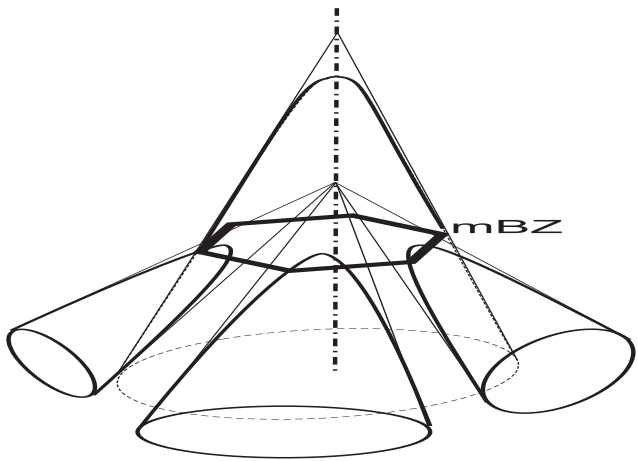}
\caption{Mini Brillouin zone (mBZ) formed from the Dirac cone
replicas in the vicinity of the primary Dirac cone.
Three of six replicas are shown. }
\end{figure}

To understand what numerical value it could correspond to, we
choose $\vec q \propto \vec K_A$ and find $\Delta \vec q_{AB} $
based on (\ref{rotate}) and expressions (\ref{sab0}, \ref{sba0})
for $\Sigma_{BA},\Sigma_{AB}$. Then the rotation angle
$\alpha=\arccos\left(\vec q_{AB}\cdot\vec q_{BA}/|\vec
q_{AB}||\vec q_{BA}|\right)\approx \pi/2$, so such a rotation
could be large enough for some points in momentum space.

Thus, the removal of the degeneracy leads to the appearance of the
hexagonal mini Brillouin zone in the vicinity of the Dirac point,
such that the corners of the cones lie on the same line, whereas
the replicas are rotated with respect to Dirac cone (Fig.~
4, 
5).
\begin{figure}[hbt] \label{fig4}
\includegraphics[width=7.5cm,height=6.6cm,angle=0]{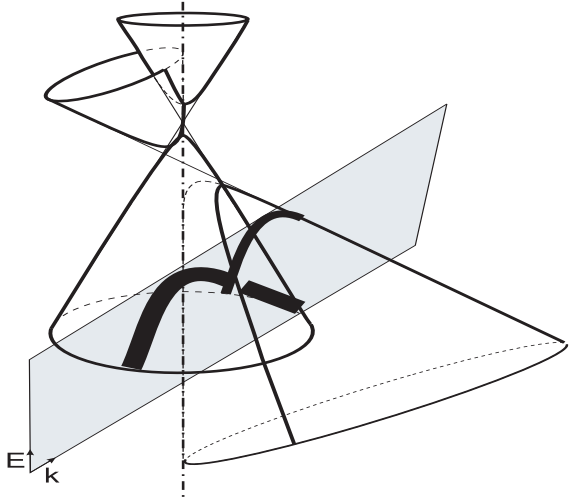}
\caption{ARPES experimental plane(shaded) cuts the valent
zone  and  its replica in the vicinity of the Dirac point $K_{A(B)}$.
 Lines at intersection mark the ARPES band (thick line) and its
 replica (less thick line). }
\end{figure}
The electron density distribution is unstable at the intersection
of the cones of the mini Brillouin bands with the Dirac cone of
the graphene Brillouin zone. Therefore, in these places the Dirac
cone and its replicas can only quasi-cross to form energy
mini-gaps  in the ARPES band as shown in Fig.~
5. Since $\Delta
_{BA} < \Delta _{AB}$, the probability of transitions for replica
photoelectrons is lower than for photoelectrons near the Dirac
cone. This fact is represented  in Fig.~
5 by lines at the
intersection of the valent zones with ARPES experimental plane
which have different intensities (shown in the figure by different
thicknesses). The more intensive line  (thick) represents the
ARPES band for photoelectrons near the Dirac cone.

Mini-gaps in the vicinity of the Dirac point has been
experimentally observed as minigaps ($<0.2$~eV) in asymmetrical
photoemission intensity of ARPES spectra for weakly interacting
graphene on iridium support
\cite{Pletikosi}.

\section{Conclusion}
To summarize, application of secondary quantized self-consistent
Dirac -- Hartree -- Fock approach to consider  electronic
properties of monolayer graphene with accounting of spin-polarized
states allows to coherently explain experimental results on energy
band minigaps and charge carrier asymmetry in graphene, propose a
description of valent and conduction zones shifts and gives a nice
theoretical estimation of electron and holes cyclotron masses
which is in very good agreement with known experimental data.


\end{document}